\documentclass[fleqn,10pt]{wlscirep}
\usepackage[utf8]{inputenc}
\usepackage[T1]{fontenc}
\usepackage{graphicx}
\graphicspath{ {images/} }
\renewcommand{\refname}{\centering  References}
\usepackage{algorithm}
\usepackage[noend]{algpseudocode}
\usepackage{authblk}
\usepackage[perpage]{footmisc}
\usepackage{setspace}
\usepackage{multirow}
\usepackage{caption}
\usepackage{hyperref}
\usepackage{doi}

\usepackage{colortbl}
\usepackage{lineno}
\usepackage{xr}
\usepackage{hyperref}
\title{Innovative Approaches in Soil Carbon Sequestration Modelling for Better Prediction with Limited Data}

\author[1,2,3,4,*]{Mohammad Javad Davoudabadi}
\author[4]{Daniel Pagendam}
\author[1,2,3]{Christopher Drovandi}
\author[5]{Jeff Baldock} 
\author[1,2,3]{Gentry White}

\affil[1]{School of Mathematical Sciences, Queensland University of Technology, Australia;}
\affil[2]{Australian Research Council Centre of Excellence for Mathematical \& Statistical Frontiers (ACEMS);}
\affil[3]{QUT Centre for Data Science, Queensland University of Technology, Australia;}
\affil[4]{CSIRO Data61, GPO Box 2583, Brisbane, QLD 4001, Australia;}
\affil[5]{CSIRO Agriculture \& Food, Glen Osmond, South Australia, Australia;}

\affil[*]{mohammadjavad.davoudabadi@hdr.qut.edu.au}
\date{}

\begin{abstract}
Soil carbon accounting and prediction play a key role in building decision support systems for land managers selling carbon credits, in the spirit of the Paris and Kyoto protocol agreements. Land managers typically rely on computationally complex models fit using sparse datasets to make these accounts and predictions. The model complexity and sparsity of the data can lead to over-fitting, leading to inaccurate results when making predictions with new data. Modellers address over-fitting by simplifying their models and reducing the number of parameters, and in the current context this could involve neglecting some soil organic carbon (SOC) components.  In this study, we introduce two novel SOC models and a new RothC-like model and investigate how the SOC components and complexity of the SOC models affect the SOC prediction in the presence of small and sparse time series data. We develop model selection methods that can identify the soil carbon model with the best predictive performance, in light of the available data. Through this analysis we reveal that commonly used complex soil carbon models can over-fit in the presence of sparse time series data, and our simpler models can produce more accurate predictions. The published version of this study is available in Scientific Reports  (\href{https://www.nature.com/articles/s41598-024-53516-z/<10.1038/s41598-024-53516-z>}{doi:<10.1038/s41598-024-53516-z>}).

\end{abstract}

\begin{document}

\maketitle


\section{Introduction}
Large-scale carbon emission from soil, one of the planet's major carbon reservoirs, into the atmosphere has deleterious impacts on global climate change, soil quality, and crop productivity \cite{adams2011managing, shi2020age}. Soil organic carbon (SOC) could be used as a significant global sink for atmospheric carbon through land-management practices, helping to reduce the atmospheric concentration of greenhouse gases and improving agricultural productivity. 

International bodies and agreements such as the Intergovernmental Panel on Climate Change (IPCC) and the Paris and Kyoto Protocol agreements mitigate global warming by assessing the science related to climate change and reduce greenhouse gas emissions, especially $CO_2$. These agreements adopted systems of carbon accounting and trading markets. A part of these carbon markets (tracking and trading) is related to selling carbon credits by farmers, organisations certifying the credits, or providing government support for the scheme, and land-holders who apply land-management practices to sequester carbon and track the change of soil carbon sequestration in their farmlands. They usually have small datasets for tracking the changes in soil carbon as SOC sampling is time-consuming and costly.

Models can quantify changes in soil carbon stocks where there is accurate understanding of processes that govern soil carbon turnover and sequestration. Such models can also help develop a deeper understanding of the sequestration process and forecast future changes and trends in SOC. Researchers have developed computer-simulation models such as RothC \cite{jenkinson1987, jenkinson1990turnover}, and Century \cite{parton1988} to help make inferences about trends in carbon stocks using time series of measurements collected over many years. For example, to improve the accounting of field emissions in the carbon footprint of agricultural products, Peter et al.\cite{peter2016improving} assess the change of SOC based on simulations with the RothC model in one of the IPCC methodological approaches (Tier 3) and compare it with other default IPCC methods. Clifford et al.\cite{clifford} developed a statistical soil carbon model to estimate and forecast the amount of carbon sequestered on farmland.

All models have their limitations and it is commonplace for modellers to make modifications that better suit specific scenarios of interest. For instance, Farina et al. \cite{farina2013modification} modified the RothC model with the aim of improving the prediction of soil carbon dynamics in semi-arid regions.  At their core, models such as RothC partition the total SOC mass into specific pools. These pools are decomposable plant material (DPM), resistant plant matter (RPM), humified organic matter (HUM), microbial biomass (BIO), and inert organic matter (IOM) \cite{adams2011managing, capon2010soil}. Modellers are, however, free to explore alternative means of partitioning soil carbon to suit different objectives.

The vast majority of SOC models are deterministic, yielding a single possible trajectory of soil carbon dynamics for a given set of parameters and an initial condition.  On the other hand, statistical SOC models can yield ensembles of possible soil carbon trajectories.  One of the main advantages of a statistical SOC model over deterministic SOC models such as RothC is introducing this randomness and providing a probabilistic method for quantifying uncertainty around model outputs. Uncertainties in SOC models arise in many ways such as around the parameters, model inputs, dynamics, and subsequently model predictions. Statistical models help to quantify uncertainties in a SOC model by modelling the different sources of randomness. Research using statistical models and sensitivity analysis (running models for different sets of parameter values) attempts to quantify uncertainties in soil carbon model outputs \cite{jones, koo, post, juston, paul, stamati}.  Clifford et al. \cite{clifford} quantified uncertainties in model inputs,  dynamics, and uncertainties in model parameters for a one pool soil carbon in a comprehensive manner using a physical-statistical model for carbon dynamics within a framework known as Bayesian hierarchical modelling (BHM). The statistical methods used by  Clifford et al. \cite{clifford} can be computationally burdensome, especially for more complex models such as some of the models we consider in this study. In addition, differences between the various soil carbon pools (DPM, RPM, HUM, BIO and IOM) are ignored in  Clifford et al. \cite{clifford}.  Gurung et al. \cite{gurung2020bayesian} identify the most important DayCent model parameters through a global sensitivity analysis for parameterization and implement a Bayesian approach using the sampling importance resampling method to calibrate the model and produce posterior distributions for the most sensitive parameters.

Microbial biomass carbon (MBC) is an important labile soil carbon fraction and the most active component of the SOC, regulating bio-geochemical processes in terrestrial ecosystems \cite{paul1996soil}.  Consequently, this has drawn the attention of modellers when considering how the MBC should be treated and how it should interact with other pools of carbon. The importance of MBC in soil carbon decomposition has led to the development of a number of microbially-explicit SOC models in recent years \cite{luo2016toward, blagodatsky2010model, frey2013temperature, moorhead2006theoretical, riley2014long}. Several microbial models with a similar basic structure and key bio-geochemical processes have been developed to simulate warming effects on soil organic matter (SOM) decomposition \cite{allison2010soil, german2012m, wang2013development}. These models differ in model complexity and reference temperature and there have been few efforts to compare model structures. For example,  Li et al.  \cite{li2014soil} have compared these models to address this question of how microbial model predictions change with increasing model complexity, and whether these predictions differ fundamentally from models with a conventional structure. More recent studies consider the interactions of microbes in a microbially-based SOC model (SOMic version 1.0) \cite{woolf2019microbial}. Other studies compare the fit of linear and non-linear soil bio-geochemical models (SBMs) using data assimilation with soil respiration data sourced from a meta-analysis of soil warming studies \cite{bg-17-4043-2020}.

In this study, we explore the effect of relaxing some of the bio-geochemical realism of models such as RothC with respect to predicting soil carbon stocks. Bio-geochemical refers to the degree to which a model accurately represents the biological, geological, and chemical processes that govern the cycling of carbon in soil ecosystems.  Our focus is using these models with the temporally sparse datasets typically available for assessing trends in soil carbon on farms, making use of two datasets from Tarlee in South Australia and Brigalow in Queensland, Australia \cite{clifford, skjemstad2004calibration}. These two sites are in different climatic regions, and it shows we can apply our approaches to a range of climatic regions. A pertinent scientific question is whether multi-pool models such as RothC are too complex relative to the limited data that is often available to fit them on a specific parcel of land.  Therefore, we attempt to understand how model predictive performance varies when we amalgamate some of these conceptual pools in the underlying process dynamics.   Specifically, we consider: (i) a single pool model considering soil carbon as a homogeneous pool that can decay and release carbon into the atmosphere \cite{clifford}; (ii) a two-pool model in which we consider a single homogeneous pool of decomposable SOC and an IOM pool that does not decompose;  (iii) a three-pool model which considers two pools of decomposable SOC (one of them represents the biological pool) and the IOM pool; and (iv) a five-pool model considering all pools mentioned above that are present in RothC. The two and three-pool models are novel soil carbon models that we introduce in this study. Also, the five-pool model used herein is somewhat novel in terms of the statistical modelling framework it is embedded in and its simplification in terms of time-step and reduced set of parameters compared to RothC.

Our modelling framework predicts changes in soil carbon stocks and accounts for epistemic uncertainty, uncertainty in the bio-geochemical process dynamics, in a statistically defensible manner. This is particularly important in the present context. We explore structural differences in the systems of equations used to describe soil carbon process dynamics which is one of the major differences between our statistical approach and that used in the simpler regression studies (e.g.  Xie et al. \cite{bg-17-4043-2020}). We develop a state-space modelling framework used for a one-pool model by \cite{DAVOUDABADI,clifford} to the two, three, and RothC-like five-pool models. We develop a Bayesian model selection method known as leave-future-out cross-validation (LFO-CV) \cite{burkner2020approximate} to choose, for a given dataset, the best soil carbon model in terms of its out-of-sample predictive accuracy. Our approach optimally adapts to the data at hand. Fitting overly complex soil carbon models might increase the uncertainty of predictions in the presence of sparse data, and it is important when making predictions about soil carbon stocks; otherwise, a land-owner might unwittingly enter into a contract to sequester carbon that has a higher risk than anticipated. Conversely, when data are sufficiently informative, our approach supports more complexity. In addition, we explore the effect of microbes and inert organic matter on the carbon cycle decomposition by adding microbial biomass and IOM pools in the one-pool model to answer this question that by adding these pools whether we obtain better soil carbon prediction than the one-pool model in  Clifford et al. \cite{clifford}. Although there are a number of studies in the literature that consider the impact of microbial biomass on soil carbon sequestration and how this affects modelling \cite{woolf2019microbial, luo2016toward, blagodatsky2010model,huang2019variation}, our process of modelling the dynamics of microbial biomass in the SOC model, along with applying advanced Bayesian methods to estimate its model parameters, are the main differences between our study and aforementioned papers.


We organise the rest of the paper as follows. The datasets used in this study are described in Section \ref{DataBackground}. We introduce our model framework and the LFO-CV criterion in Section \ref{Methods}. In Section \ref{ModelStructure} the structure of the models is described. In Section \ref{results}, we compare the models based on their out-of-sample predictive accuracy and quantify the uncertainty of our estimate. Section \ref{SectionConclusion} presents a discussion of this study and our results.

\section{Background and Description of Datasets}\label{DataBackground}
Our model selection method is motivated by two datasets that are collected from two locations in Australia. The details of these sites are presented in the following. 
\subsection{Tarlee Dataset}
An agricultural research experiment site known as Tarlee situated $80$ km north of Adelaide, South Australia was established in $1977$ to examine the impact of management
practices on agricultural productivity as a long-term field experiment \cite{datasetCSIRO}. The soil of this site is classified as a hard-setting red-brown earth with sandy loam texture. This site has a Mediterranean climate and is dominated by winter rainfall with an average of $355$ mm from April to October \cite{clifford, luo, skjemstad2004calibration}. Soil properties of that site were monitored over a 20-year period in three fields under different management practices, and soil samples covering the entire top $30$ cm of the profile were obtained for the years 1979, 1985, and 1996 from all 3 rotations. Table \ref{ManagementTreatments} presents the time period of management treatments that were implemented in three trial fields in Tarlee.

\begin{table}[ht]
\begin{center}
\begin{tabular}{|c|c|c|c|}
\hline
 \cellcolor{gray!60} Management treatments   & \cellcolor{gray!60} Field 1 & \cellcolor{gray!60} Field 2 & \cellcolor{gray!60} Field 3 \\
\hline    
    Wheat for grain & (1979 - 1987) and  & - & - \\ & (1990 - 1996) &  & \\ 
\hline     
    Wheat for hay & 1988 and 1989 & 1989 & - \\
\hline    
    Fallow & 1997 & 1997 & 1997 \\
\hline    
    Wheat for grain and fallow & - & (1979 - 1988) and & - \\ &  &  (1990 - 1996) & \\
\hline 
    Wheat and pasture & - & - & (1979 - 1987) \\
\hline 
    Wheat and pasture for hay & - & - & 1988 and 1989 \\
\hline 
    Wheat for grain and pasture & - & - & (1990 - 1996) \\
\hline

\end{tabular}
\caption{\label{ManagementTreatments} The duration of management treatments in three fields in Tarlee.}
\end{center}
\end{table}

\subsection{Brigalow Dataset}
Brigalow is a research station in Queensland, Australia. This site is situated in a semi-arid, and subtropical climate, and consists of three forested catchments of 12-17 ha \cite{skjemstad2004calibration}. Three monitoring sites were established within each of the catchments in recognition of three soil types (a duplex soil and two clays). One catchment was planted to wheat and occasional sorghum and the other to buffel pasture and the last one was left under native Brigalow forest. At this site, on one catchment, after clearing land under Brigalow (\textit{Acacia harpophylla}) in $1982$, continuous wheat with some sorghum was established over a 18-year period. Samples were collected from the field in two distinct categories: surface samples, acquired from a depth of 0-10 cm, and profile samples, retrieved down to a depth of 200 cm. In the profile category, samples were taken at three specific intervals within the upper layers: 0-10 cm, 10-20 cm, and 20-30 cm. Table \ref{BrigalowDataTable} shows the duration of management practices in Brigalow.

\begin{table}[ht]
\begin{center}
\begin{tabular}{|c|c|c|c|}
\hline
 \cellcolor{gray!60} Management treatments   & \cellcolor{gray!60} Soil type 1 & \cellcolor{gray!60} Soil type 2 & \cellcolor{gray!60} Soil type 3 \\
\hline    
    Cleared & 1982  & 1982 & 1982 \\ 
\hline 
    Wheat for grain & (1985 - 1992) and  & (1985 - 1992) & (1985 - 1992) \\ & (1994, 1996, 1998) & (1994, 1996, 1998) & (1994, 1996, 1998) \\ 
\hline     
    Sorghum for grain & 1984, 1995, & 1984, 1995, & 1984, 1995,  \\ &  1997 and 1999 &  1997 and 1999 & 1997 and 1999 \\
\hline    
    Fallow & 1983 and 1993 & 1983 and 1993 & 1983 and 1993 \\
\hline

\end{tabular}
\caption{\label{BrigalowDataTable} The duration of management treatments in Brigalow.}
\end{center}
\end{table}

\section{Methods}\label{Methods}
\subsection{Soil Carbon Model}\label{SoilCarbonModel}
We can consider uncertainties in a dynamical SOC model as arising from three sources: errors in the observations, randomness or uncertainty inherent in the underlying physical processes, and uncertainties in model parameters \cite{clifford}. These uncertainties are modelled through the observation model $p(\mathbf{Y} | \mathbf{X}, \boldsymbol{\theta})$, the process model $p(\mathbf{X}| \boldsymbol{\theta})$, and the prior $p(\boldsymbol{\theta})$. Here $\boldsymbol{\theta}$, $\mathbf{Y}$, and $\mathbf{X}$ denote unknown parameters, observations, and unobserved state process, respectively. Furthermore, the probability density function of the enclosed random variable, and the conditional probability density function given the event $E$ are denoted by $p(.)$, and $p(.|E)$, respectively. For example, the mass of SOC, $X_C$, is one of the elements of $\mathbf{X}$, or the measured value of total SOC, $Y_{TOC}$, is one of the elements of $\mathbf{Y}$, furthermore, the decay rate of total SOC, $K_C$, is an example of a model parameter in a soil carbon model. 

These three models form a  hierarchical framework known as a Bayesian Hierarchical Model (BHM). The top level of the hierarchy contains the observation model which includes noisy observational data that depend on the state variables. This model is followed by the process model, located at the second level. At this level, latent state variables, which cannot be measured directly but can be estimated based on measurement data that depend on the latent state variables, are modelled. These two models typically rely on some unknown parameters. The third level underneath these two levels contains the parameter model
\cite{allenby2006hierarchical,berliner1996hierarchical,cressie2015statistics}. A BHM is represented mathematically as follows:
\begin{align}\label{BHM hierarcy}
    p(\mathbf{Y}, \mathbf{X}, \boldsymbol{\theta}) = p(\mathbf{Y},  \mathbf{X} | \boldsymbol{\theta}) p(\boldsymbol{\theta}) = p(\mathbf{Y} | \mathbf{X}, \boldsymbol{\theta}) p(\mathbf{X} | \boldsymbol{\theta}) p(\boldsymbol{\theta}).
\end{align}

\noindent Note that the joint distribution $p(\mathbf{Y}, \mathbf{X}, \boldsymbol{\theta})$ captures all the uncertainty in the model. The advantage of analysing a model within the BHM framework is that it incorporates prior knowledge related to the parameters into the analysis by updating the distributions of these parameters with observed data. The latent state of the SOC, $\mathbf{X}$, evolves as a  dynamical process and given noisy, sparse data.  Inferences about soil carbon dynamics, parameters, and functions of them can be made through the posterior distribution $p(\mathbf{X}, \boldsymbol{\theta}|\mathbf{Y})$. We can write the posterior distribution based on (\ref{BHM hierarcy}) as follows:
\begin{align}\label{Bayes formula}
   p(\mathbf{X}, \boldsymbol{\theta}|\mathbf{Y}) = \frac{p(\mathbf{Y} | \mathbf{X}, \boldsymbol{\theta}) p(\mathbf{X} | \boldsymbol{\theta}) p(\boldsymbol{\theta})}{p(\mathbf{Y})}
\end{align}
where $p(\mathbf{Y})$ depends only on data and may be difficult to calculate analytically or numerically, thus the posterior itself may
be difficult to evaluate. Fortunately, one can draw samples from the posterior if it is not analytically tractable. 

As in other recent statistical analyses \cite{DAVOUDABADI, clifford} we use a state-space modelling framework, the first and second levels of the BHM, to predict changes in soil carbon stocks. State-space models are more challenging to fit in practice than simpler regression models used in   \cite{bg-17-4043-2020} because they acknowledge uncertainty in the latent process dynamics. The prior information of the parameter model in the third level of the BHM is described in the following.

\subsection{Prior Information}\label{Prior_Info}
As mentioned earlier, the process model and the observation model typically depend on unknown parameters, and the parameter model captures the uncertainty around these parameters. A Bayesian approach for model fitting is applied to quantify the uncertainty in parameters and predictions. This approach places a prior distribution on the unknown parameter vector $\boldsymbol{\theta}$, which is the advantage of using the Bayesian analysis since we implement our prior knowledge of parameters as part of the inferential process. 

In general, the prior knowledge about parameters includes three categories: informative, weakly informative, and uninformative priors. When we have a small dataset or the dataset is sparse, the prior distribution becomes more influential and informative priors can become more useful. In this study, we obtain priors from previous studies \cite{clifford, DAVOUDABADI, skjemstad2004calibration} and expert
opinion. The model parameters and their prior probability density functions are listed in Supplementary Tables S2 
and S3 
(Section B 
of the supplementary material).

\subsection{Posterior Distribution Inference}\label{Methodology}
To estimate the changes in SOC over time as a result of the various management practices, and to estimate the parameters driving the sequestration of carbon, we sample from the posterior distribution $p(X_{TOC}, \boldsymbol{\theta} | \mathbf{Y})$, where $X_{TOC}$ is the mass of total SOC. To this end, we draw samples from the posterior distribution $p(\mathbf{X}, \boldsymbol{\theta}|\mathbf{Y})$ in (\ref{Bayes formula}) which can be decomposed into two components $p(\mathbf{X}|\boldsymbol{\theta}, \mathbf{Y})p(\boldsymbol{\theta} | \mathbf{Y})$ and we preserve the components related to the SOC process $X_{TOC}$ and its parameters $\boldsymbol{\theta}$. Davoudabadi et al.  \cite{DAVOUDABADI} used advanced Bayesian methods, e.g. correlated pseudo-marginal (CPM) method and the Rao-Blackwellised particle filters (RBPF) for state-space models, to reduce the computational cost of estimating uncertainties in the one-pool model presented by \cite{clifford}. The CPM method, one of several particle Markov chain Monte Carlo (PMCMC)
methods, is applied to the model to draw samples from $p(\boldsymbol{\theta} | \mathbf{Y})$ as the resulting likelihood is not tractable \cite{deligiannidis2018correlated, DAVOUDABADI}. The CPM method in Davoudabadi et al. \cite{DAVOUDABADI} outperforms other state of the art PMCMC methods in terms of computation time. The advantage of using this method is that it reduces the computational cost of estimating intractable likelihoods by correlating the estimators of the likelihoods in the acceptance ratio of its algorithm. Algorithm S3 
in Section C.3 
of the supplementary material provides the  CPM algorithm. This correlation can be achieved by correlating the auxiliary random numbers used to obtain these estimators; see Davoudabadi et al. \cite{DAVOUDABADI} and Deligiannidis et al.\cite{deligiannidis2018correlated} for more details. To estimate the marginal likelihood of the state variables, we use the RBPF as the SOC model combines linear and non-linear sub-models. The RBPF algorithm estimates the marginal likelihood of the non-linear sub-model through bootstrap particle filter (BPF). It computes the marginal likelihood of the linear part of the model through the Kalman Filter (KF) algorithm \cite{doucet2000rao, DAVOUDABADI}. Computing the exact likelihood of the linear sub-model makes the RBPF algorithm an attractive algorithm in these scenarios as it reduces the computational cost of the estimated likelihood dramatically. See Davoudabadi et al. \cite{DAVOUDABADI} for more details about the RBPF, BPF and KF algorithms. In addition, the algorithm of the KF and BPF methods are provided in Sections C.1 
and C.2 of Supplementary Material 
, Algorithms S1 
and S2 
, respectively. The RBPF algorithm is reused to draw a sample of the state process from the posterior distribution $p(X_{TOC}|\boldsymbol{\theta}, \mathbf{Y})$. In the CPM algorithm, it is required to generate candidate parameters from appropriate proposal distributions. More precisely, a proposal distribution is a user-specified distribution that the user is free to choose and the Markov chain will converge to the desired posterior distribution if it is run for enough iterations. However, a proposal distribution can have a significant impact on the finite-time efficiency of the MCMC and the ideal case occurs when the proposal distribution is the desired posterior distribution which is typically unknown. The proposal distributions are presented in the supplementary material Section B. 

We can quantify the uncertainty of our estimate in many ways, for example, through a $95\%$ credible interval or the estimated expected value of functionals of interest. The inference about the mass of SOC added over a period of time can be achieved through the MCMC samples of the posterior distribution. We represent the posterior distribution $p(\boldsymbol{X},\boldsymbol{\theta} \vert \boldsymbol{Y})$ by $M^*$ samples $\lbrace (X^m,\boldsymbol{\theta} ^m) : m = 1,...,M^* \rbrace$ and the posterior expectation of any function $g^*(\boldsymbol{X}, \boldsymbol{\theta})$ can be estimated by these samples. 
\begin{align*}
    \mathbf{E}(g^*(\boldsymbol{X}, \boldsymbol{\theta}) \vert \boldsymbol{Y}) \approx  \frac{1}{M^*} \sum_{m=1}^{M^*} g^*(X^m,\boldsymbol{\theta} ^m).
\end{align*}
The error of the accuracy of such estimates is negligible for sufficiently large sample size $M^*$. The change in SOC to field $i$ between the first year of trial, e.g. $t=1$, and following year $t$ in a dataset is considered as follows
\begin{equation*}
    g^*(\boldsymbol{X}, \boldsymbol{\theta}) = X_{TOC(t)}^i - X_{TOC(1)}^i;
\end{equation*}
and can be estimated as follows
\begin{equation*}
   \hat{g^*}(\boldsymbol{X}, \boldsymbol{\theta}) = \mathbf{E} (X_{TOC(t)}^i - X_{TOC(1)}^i\vert \boldsymbol{Y});
\end{equation*}
where $X_{TOC(t)}^i$ is the summation of other pools, for example, in the three-pool model $X_{TOC(t)}^i$ is equal to the summation of $X_{C(t)}^i$, $X_{IOM(t)}^i$, and $X_{B(t)}^i$. 

\noindent The posterior variance, $\mathrm{var} (X_{TOC(t)}^i - X_{TOC(1)}^i\vert \boldsymbol{Y})$, is a measure of uncertainty associated with this Bayes estimate.

We assess the quality of the MCMC samples through an MCMC diagnostic known as the Gelman and Rubin's convergence diagnostic statistic \cite{gelman1992inference}. The Gelman and Rubin's convergence diagnostic statistic, $\hat{R}$, can be used to assess whether the MCMC samples have ``mixed'' sufficiently, effectively sampling from the probability distribution, and have reached a stationary distribution \cite{gelman1992inference}. Gelman and Rubin's convergence diagnostic compares samples from multiple chains to assess whether the output from each chain is sufficiently similar to the others. The output from each chain is indistinguishable from the others when the scale reduction factor estimated from the sampling is less than 1.2 \cite{brooks1998general}. 

Before estimating model parameters and conducting inference with a model, it is essential to validate our model to establish its suitability for estimating changes in soil carbon stocks. In the next section, we introduce our method for selecting between competing soil carbon models, focusing on predictive accuracy.

\subsection{Model Evaluation}\label{ModelSelectionSection}
One way to evaluate a model or compare different models is to measure predictive accuracy \cite{gelman2014understanding}. As our models depend on time, for model comparison and selection, we apply leave-future-out cross-validation (LFO-CV) that refits a model to different subsets of the data \cite{burkner2020approximate}. The LFO-CV is a fully Bayesian metric in that it uses the entire posterior distribution. This method is the approach used to compare the model’s predictive accuracy for the four SOC models listed in Section \ref{ModelStructure}. 

Let $Y_{1:T}$ be a time series of observations and let $L$ be the minimum number of observations from the series that we will require before making predictions for future data. To make reasonable predictions for $Y_{i+1}$ based on $Y_{1:i}$, $i$ should be large enough so that we can learn enough about the time series to predict
future observations, otherwise, it may not be possible to make reasonable predictions. The choice of $L$ depends on the application and how informative the data are, therefore, it may be vary from one dataset to another \cite{burkner2020approximate}. We would like to compute the predictive densities $p(\Tilde{Y}_{t+1}|Y_{1:t})$ for each $t \in \{L,...,T-1\}$ where $\Tilde{Y}_{t+1}$ is a future vector of observed data. The expected log pointwise predictive density (ELPD) can be used as a global measure of predictive accuracy, which is
\begin{align}\label{ELPD}
    \mbox{ELPD}= \log \prod _{t=L}^{T-1}  \mathbf{E}_{\theta|Y_{1:t}}(p(\Tilde{Y}_{t+1}|Y_{1:t},\theta)) = \sum _{t=L}^{T-1} \log \int p(\Tilde{Y}_{t+1}|Y_{1:t},\theta)p(\theta|Y_{1:t})~d\theta.
\end{align}
In practice, the integral in (\ref{ELPD}) is intractable, however we can approximate it through Monte-Carlo methods \cite{burkner2020approximate}. To estimate $ p(\Tilde{Y}_{t+1}|Y_{1:t})$, we draw samples $(\theta_{1:t}^1,..., \theta_{1:t}^S)$ from the posterior distribution $p(\theta|Y_{1:t})$ for $t \in \{1,...,\gamma\}$ where $\gamma \in \{L,...,T-1\}$ using the particle MCMC method described in Section \ref{Methodology} and estimate the predictive density for $\Tilde{Y}_{L+1:T}$ as follows
\begin{align}\label{PridictDensity}
    p(\Tilde{Y}_{t+1}|Y_{1:t}) \approx \frac{1}{S}\sum_{s=1}^S p(\Tilde{Y}_{t+1}|Y_{1:t}, \theta_{1:t}^s).
\end{align}
When our model is a state-space model, we need to consider the state variables as part of the parameter space and estimate them through the particle filter methods to apply the LFO-CV. The reason for selecting ELPD instead of other global measures of accuracy such as the root mean squared error (RMSE) is that it evaluates a distribution to provide a measure of out-of-sample predictive performance rather than evaluating a point estimate like the mean or median, which we see as favourable from a Bayesian perspective \cite{vehtari2012survey, burkner2020approximate}.

\section{Model Structure}\label{ModelStructure}
\begin{figure}[ht]
\begin{center}
 \includegraphics[scale=0.6]{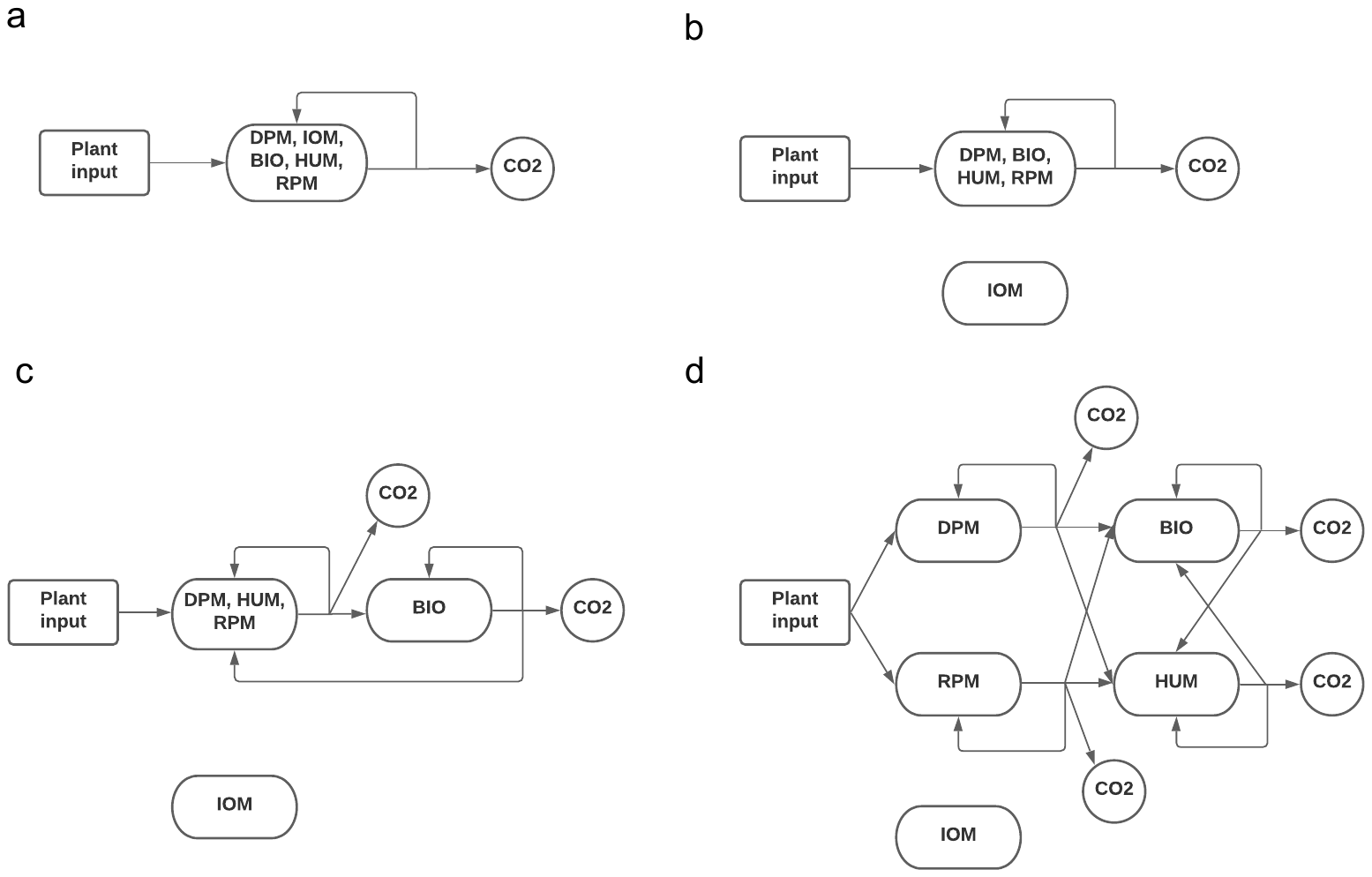}
 \vspace*{-15mm}
 \caption{Graphical representation of the carbon emission in the a) one-pool model, b) two-pool model, c) three-pool model, and d) five-pool model. The five pools from RothC have been amalgamated into a single homogeneous soil carbon pool in the one-pool model. The DPM, BIO, HUM and RPM pools are amalgamated and treated as a single homogeneous pool in the two-pool model, and the DPM, HUM and RPM pools are amalgamated and treated as a single homogeneous pool in the three-pool model.} 
 \label{figOnePool}
\end{center}
\end{figure}

The total SOC consists of different components defined by their origin and their decay rate. These components originate from living organisms known as biotic material or non-living (abiotic) material \cite{adams2011managing,lal2010managing}. Based on the RothC model, the components of the total SOC include DPM, RPM, HUM, BIO, IOM \cite{adams2011managing,capon2010soil}. The one-pool model in Clifford et al.\cite{clifford} considered all components mentioned above as a single pool. The process model of the one-pool model is a combination of linear and non-linear sub-models. The details of the process and the observation models of these sub-models are shown in the supplementary material Sections D.1 
and D.2 
, respectively. Figure \ref{figOnePool}a graphically represents the carbon emission process in the one-pool model. Based on Figure \ref{figOnePool}a, a fraction of carbon decays is emitted into the atmosphere as $CO_2$ and the rest remains in the pool.

In the two-pool model, we consider the IOM pool as a second pool that is resistant to chemical and biological reactions and encompasses charcoal or charred material  \cite{capon2010soil}. The IOM fraction is not subject to biological transformation and is thus constant \cite{falloon2000important}. As the IOM fraction is constant, its process model at time $t$ is a constant value and should be estimated. The process and the observation models of the two-pool model are presented respectively in Sections  E.1 
and E.2 
. Figure \ref{figOnePool}b shows the graphical representation of the two-pool model.

The three-pool model considers the IOM and BIO as separate pools with a main pool of decomposable carbon which is an amalgamation of DPM, RPM, and HUM pools. Soil carbon decomposes from the decomposable carbon pool, and fractions are either transferred to the BIO pool or lost to the atmosphere as $CO_2$. Carbon present in the BIO pool that decomposes is either lost to the atmosphere as $CO_2$, re-assimilated as biological mass or transferred to the main soil carbon pool. Figure \ref{figOnePool}c shows the diagram of the carbon emission in the three-pool model. The process and observation models of the three-pool model are presented in detail in Sections F.1 
and F.2 
of the supplementary material, respectively. It is noteworthy to mention that the size of the microbial pool encompasses a small fraction of the total organic carbon, e.g. $5\%$ of the TOC, based on expert knowledge. We implement this constraint by rejecting BIO state trajectories that exceed $5\%$ of the TOC in the Markov chain Monte Carlo (MCMC) algorithm.

The RothC model, consisting of five conceptual pools, is the standard soil carbon used in many studies and is considered a reasonable representation of the physical sub-species of carbon in the soil. In the models presented so far, we have considered the pools to be either one of the RothC pools or an amalgamation of the five RothC pools. In the five-pool model presented here, we now retain the structure presented in the RothC model without any amalgamation.  

\noindent In the five-pool model, plant material is split between two conceptual pools: DPM and RPM.  Decomposition of carbon from these two pools either leaves the system as $CO_2$ or is transformed to carbon in the BIO and HUM pools.  Carbon from the BIO and HUM pools that decomposes can either be lost to the atmosphere as $CO_2$, or transformed to carbon in the BIO or HUM pools. The process and observation models of the carbon transfer in the five-pool model are presented mathematically in detail in Section G 
of the supplementary material. The five-pool model is depicted in Figure \ref{figOnePool}d.

  

\section{Results}\label{results}
\subsection{Comparing Models}\label{ComparingModels}
\begin{figure}[ht]
    \centering
    \includegraphics[width=16cm, height=10.2cm]{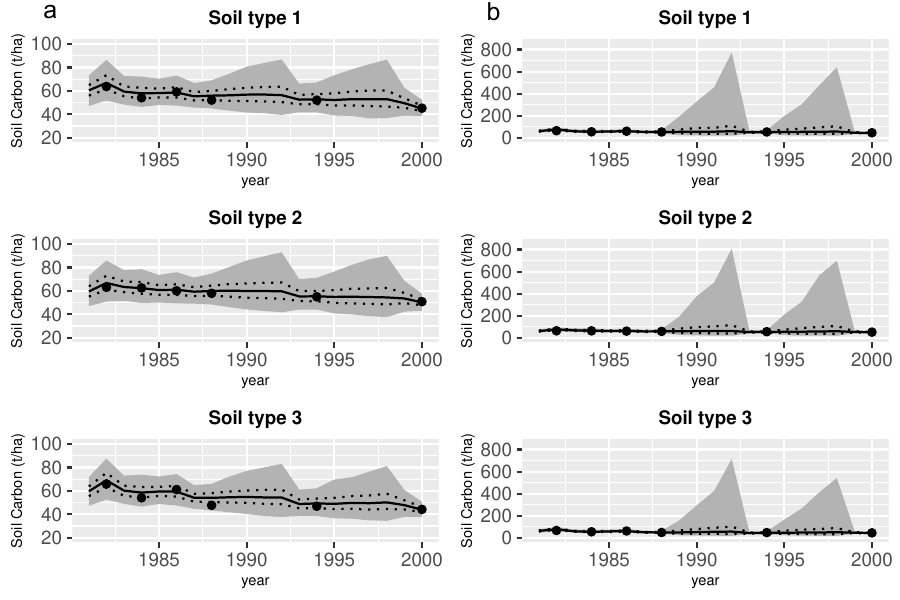}
    \vspace*{-0.5cm}
    \caption{Soil organic carbon (SOC) dynamics of the Brigalow dataset based on a) the three-pool model and b) the five-pool model.  The gray shaded part is the area between the $2.5^{th}$ and the $97.5^{th}$ percentiles for the SOC process gained by the three and five-pool models. The $25^{th}$ and the $75^{th}$ percentiles for the SOC process are indicated by the dashed lines. The $50^{th}$ percentile is shown by the solid line and the measured SOC values are indicated by filled dots.}
    \label{BrigalowTrajectories}
\end{figure}

We worked with four MCMC chains,  each initialised with a randomly sampled parameter vector, in the Correlated Pseudo-marginal Method (CPM) method for estimating the predictive density (\ref{PridictDensity}). We ran each chain for 200,000 iterations discarding the first 80,000 as burn-in. We thinned these chains, choosing every $30^{th}$ sample of the MCMC samples to estimate (\ref{PridictDensity}), therefore, $S$ in equation (\ref{PridictDensity}) was equal to 4,000. The minimum numbers of observations, $L$, used for making predictions for future data in the Tarlee and Brigalow datasets were $12$ and $13$, respectively. The estimated expected log pointwise predictive density (ELPD) of the one, two, three, and five-pool models applied on the Tarlee dataset were $-53.02$, $-40.55$, $-34.79$, and $-37$, respectively. The estimated ELPD of those models applied on the Brigalow dataset were $-36.89$, $-36.88$, $-36.48$, and $-49.57$, respectively. Based on these results (supplementary material Tables S13 
and S14) 
, the three-pool model outperformed the other models in the sense of yielding the best LFO predictive ability for both the Brigalow and Tarlee datasets. This three-pool model included an inert carbon pool and two decomposable pools that were conceptually equivalent to a biological pool (the decomposers) and a decomposable material pool, an amalgamation of DPM, RPM, and HUM pools. For Tarlee, the five-pool RothC-like model had the next best ELPD, but in Brigalow, the five-pool model exhibited the worst ELPD of the four models studied. The performances of the three and five-pool models in estimating the trajectories of the SOC dynamics of the Brigalow dataset are highlighted visually in Figure \ref{BrigalowTrajectories}a and \ref{BrigalowTrajectories}b, respectively. As shown in Figure \ref{BrigalowTrajectories}b, the five-pool model increased uncertainty in the soil carbon dynamics, especially during the sparse periods, typified by wide $95\%$ credible intervals. The significant variability in these regions stems from our practice of simulating input state values, such as the total mass of wheat dry matter ($X_W$), during each iteration of the particle filter algorithm and subsequently aggregating them. However, when there is no observation available for comparing these simulated values, it introduces additional variability in the trajectory of the state variables. Hence, when an observation ($Y_{(t)}$) is present, the level of uncertainty is notably lower compared to other scenarios.

Setting aside the five-pool model and focusing on the one, two, and three-pool models, we see that amongst these three models, the ranking from best to worst is three-pool, two-pool, and one-pool for both study sites. We cannot say with full confidence the three-pool model is the best model for the Brigalow dataset compared to the one and two-pool models as there is not much difference between their estimated ELPDs acknowledging the Monte Carlo errors. Nevertheless, we select it as the best model for the Brigalow dataset since the three-pool model has the largest ELPD.



\subsection{Uncertainty Quantification} \label{inference}
\begin{figure}[ht]
    \centering
    \includegraphics[width=16cm, height=10.2cm]{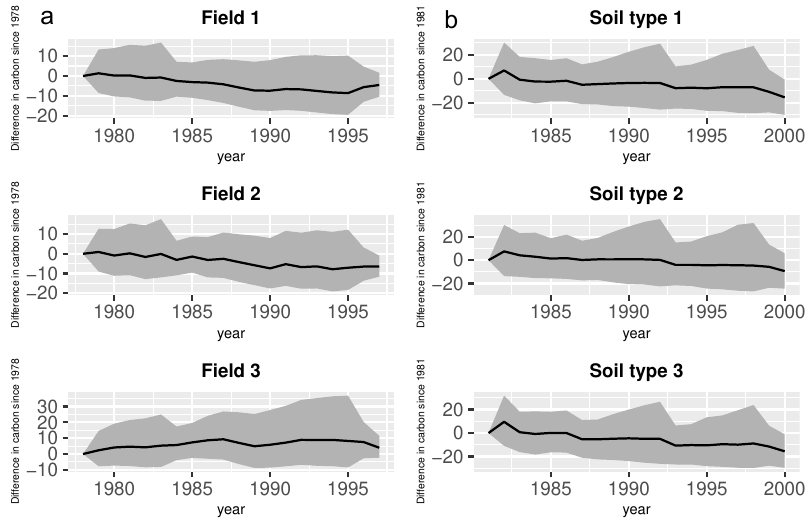}
    \vspace*{-0.1cm}
    \caption{The expected difference of the SOC in each year from 1978 and 1981 in the a) Tarlee and b) Brigalow datasets, respectively, estimated based on the three-pool model. The change of the SOC stock in each field/soil type is indicated by solid line, and the gray shaded part is the area between the $2.5^{th}$ and the $97.5^{th}$ percentiles for the SOC process.}
    \label{DifferenceInCarbonSince1978}
\end{figure}

The average of the SOC change between 1978 and 1997 in fields 1, 2, and 3 in the Tarlee trial based on the three-pool model were $-3.81$, $-3.47$, and $7.12$, respectively (Figure \ref{DifferenceInCarbonSince1978}a). Here the negative values denote that the first two fields were expected to lose carbon over the 20-year period. The management strategies that are used in fields 1, 2,
and 3 are ``Wheat-Wheat”, ``Wheat-Fallow”, and ``Wheat-Pasture”,
respectively. This average for three soil types of the Brigalow dataset, based on the three-pool model, between 1981 and 2000 were $-4.37$, $-0.43$, and $-5.13$, respectively (Figure \ref{DifferenceInCarbonSince1978}b). The hardware use and computing time information are provided in Section J 
of the Supplementary Material. 

\begin{figure}[ht]
    \centering
    \includegraphics[width=13cm, height=10.2cm]{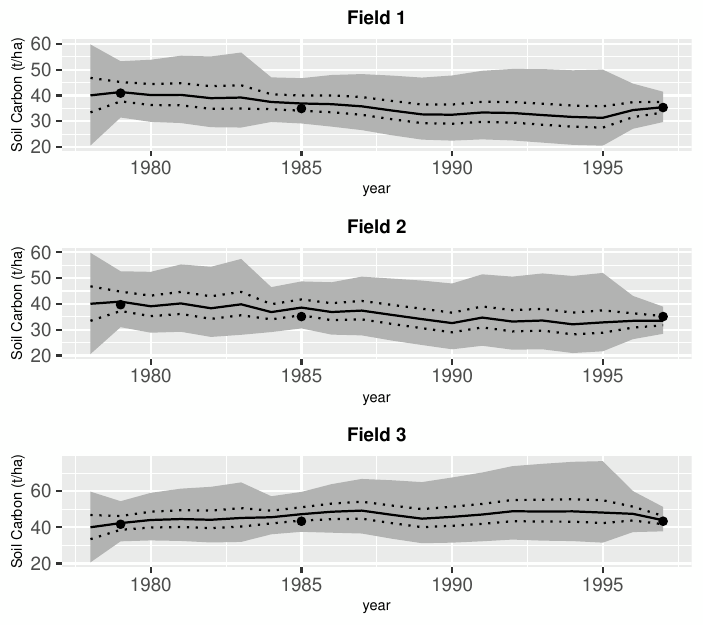}
    \vspace*{-0.5cm}
    \caption{Soil organic carbon (SOC) dynamics in the three Tarlee fields.  The gray shaded part is the area between the $2.5^{th}$ and the $97.5^{th}$ percentiles for the SOC process from the three-pool model. The $25^{th}$ and the $75^{th}$ percentiles for the SOC process are indicated by the dashed lines. The $50^{th}$ percentile is shown by the solid line.  The measured SOC values are indicated by filled dots.}
    \label{TarleeTrajectories}
\end{figure}

We can find the $95\%$ credible interval for the amount of carbon in the soil by computing the upper and lower limits of the interval which are the $97.5^{th}$ and $2.5^{th}$ percentiles of the posterior distribution, respectively. These percentiles for the SOC process of each soil type in the Brigalow trial and each Tarlee field are presented in Figures \ref{BrigalowTrajectories}a and \ref{TarleeTrajectories}, respectively. Due to the wide range of soil carbon stocks in Figure \ref{BrigalowTrajectories}(b) we also provide a separate comparison of the $50^{th}$ percentiles based on three and five-pool models for Brigalow  in Supplementary Figures S1a 
 and S1b, 
 respectively in section Supplementary Material. 

\begin{figure}[ht]
    \centering
    \includegraphics[width=15cm, height=12.2cm]{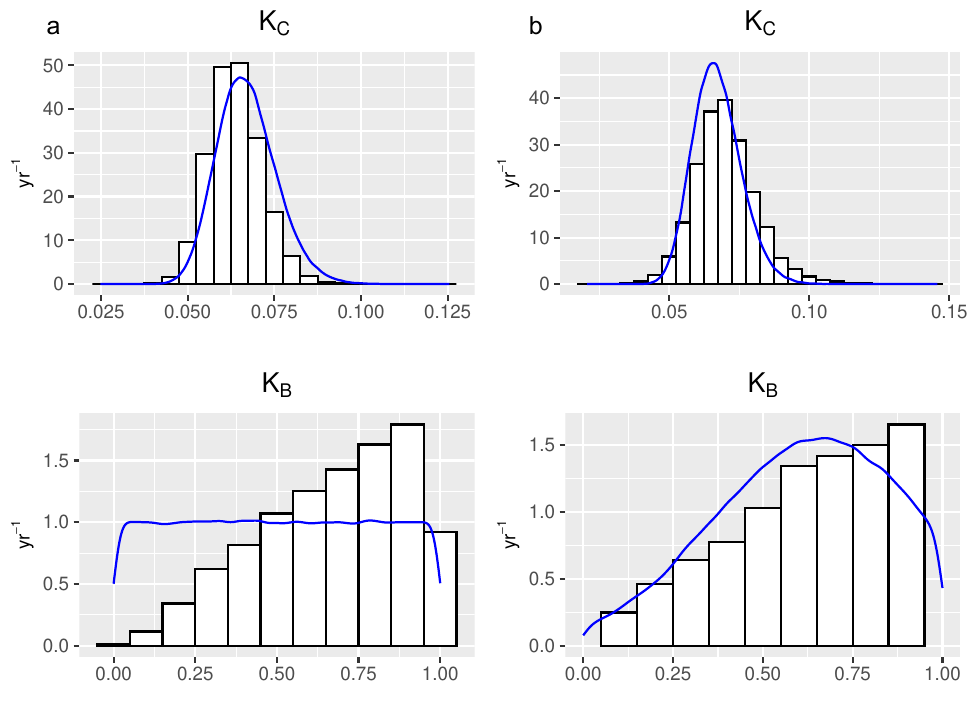}
    \vspace*{-0.5cm}
    \caption{The marginal posterior distributions (histogram) of the SOC and BIO decomposition rates, $K_C$ and $K_B$, respectively, in a) Tarlee and b) Brigalow. The histograms correspond to the three-pool model in both Brigalow and Tarlee. The blue densities are the prior distributions of the SOC and BIO decomposition rates.}
    \label{Pri_post_3Pool_TB}
\end{figure}

\begin{figure}[ht]
    \centering
    \includegraphics[width=15cm, height=12.2cm]{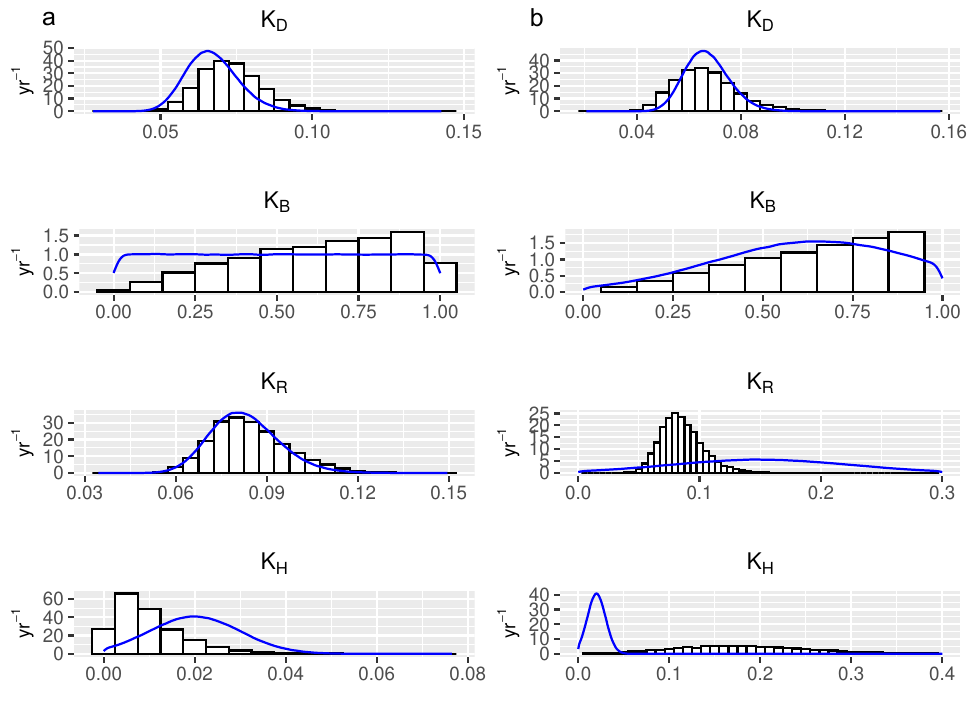}
    \vspace*{-0.5cm}
    \caption{The marginal posterior distributions (histogram) of the DPM, BIO, RPM and HUM decomposition rates $K_D$, $K_B$, $K_R$, and $K_H$, respectively, in a) Tarlee and b) Brigalow. The histograms are correspond to the five-pool BIO-K model in both Brigalow and Tarlee. The blue densities are the prior distributions of the DPM, BIO, RPM and HUM decomposition rates.}
    \label{Pri_post_5Pool_TB}
\end{figure}

As mentioned earlier in Section \ref{Prior_Info}, prior knowledge plays a significant role in the presence of small and sparse datasets. We compare the prior distributions with a histogram of the samples drawn from the posteriors of some main model parameters of the three and five-pool models that are the best and the more complex models, respectively, to highlight what we have learned about those parameters. Figures \ref{Pri_post_3Pool_TB}a and \ref{Pri_post_3Pool_TB}b show the difference between the prior and posterior of the decomposition rate of the SOC and BIO pools of the three-pool model in Tarlee and Brigalow, respectively. Also, Figures \ref{Pri_post_5Pool_TB}a and \ref{Pri_post_5Pool_TB}b show the difference between the prior and posterior of the decomposition rate of each pool of the five-pool model in Tarlee and Brigalow, respectively. Based on Figures \ref{Pri_post_3Pool_TB} and \ref{Pri_post_5Pool_TB}, it is clear that we learn quite a lot about some parameters such as $K_B$ and $K_H$, and we learn little new about other parameters, namely $K_C$ and $K_D$ as the posterior and prior are very similar.

We calculated the Gelman and Rubin's convergence diagnostics, $\hat{R}$ for the model parameters of the three-pool model of the Tarlee dataset and the one-pool model of the Brigalow dataset. They are presented in Supplementary Tables S11 
and S12, 
respectively, in Section H 
of the supplementary material. Since the values of $\hat{R}$ are less than 1.2, there is no evidence of divergence.

\newpage
\section{Discussion}\label{SectionConclusion}
In this study, we have developed three new soil carbon models and compared them with the one-pool model in Clifford et al. \cite{clifford} in the BHM framework, which allows us to think conditionally and critically about the parameters, the process, and the data that reside within a soil carbon model. To show these models are broadly applicable, we have implemented them for two datasets. 

An important motivating question behind this study is whether multi-pool state-space models based on deterministic models such as RothC are fit for making inferences on soil carbon dynamics in commonly occurring situations where soil carbon measurements are monitored infrequently.  In fitting models to two Australian datasets, we found a three-pool model (in both the cases of Tarlee and Brigalow) to have the best predictive ability of those models considered and to be better than a five-pool model, which is frequently adopted for its bio-geochemical realism.  We conclude that the detail and realism included in statistical soil carbon models should consider the volume and quality of data available for making inferences.  Indeed, this study has shown that some concessions in physical realism can lead to better predictive accuracy. This can be helpful for the IPCC, Paris agreement and Kyoto protocol's purposes, especially for national carbon accounting where datasets are sparse. 

Furthermore, we have explored the effect of microbes and inert organic matter on the carbon cycle decomposition by adding microbial biomass and IOM pools in the Tarlee model in Clifford et al. \cite{clifford}. In particular, based on the LFO-CV criterion, we have shown that the three-pool model, which includes microbial biomass and IOM pools, outperforms other models on the Tarlee and Brigalow datasets. The LFO-CV of the five-pool model is close to the three-pool model in its predictive ability for Tarlee but not for Brigalow. The reason is that the Brigalow dataset has more uninformative priors and sub-models than the Tarlee dataset. Both the Brigalow and Tarlee datasets exhibit relatively long, multi-year periods with no observation of any carbon pools, i.e. temporally sparse data. During those periods, all knowledge about the soil carbon process comes from the carbon inputs, the process dynamics and the model parameters through prior distributions. However, in the case of Brigalow, adding more pools to the model increased uncertainty in the soil carbon dynamics in each iteration of the particle filter process, causing wide variance which make it a poor predictor, typified by wide $95 \%$ credible intervals during those sparse periods. This result indicated that multi-pool models might not be as fit-for-purpose compared to some simpler models when used with sparse data over time. 

In exploring soil carbon models with reduced complexity, we chose not to investigate a four-pool model.  We could create such a model by combining the DPM and RPM components, for example.  However, we deemed a four pool model to be too similar in structure to the five pool model, therefore not providing much additional variation in model complexity.  Furthermore, our aims in this study were to explore the importance of microbe and inert organic matter pools because they are fundamentally different from other soil carbon pools (the former being constrained in its total pool size and the latter being stable over very long time scales).  The range of models used in this study provides valuable insight into whether the complexity of the RothC model is warranted when datasets are temporally sparse.

We have shown that, the three-pool model that was found to be best suited to the Brigalow and Tarlee datasets in this study can be used to obtain good fits to observational data and can be used to estimate with uncertainty the net gain or loss of carbon overtime at each study site.

Since both datasets used in this study are not large, we have used the LFO-CV criterion for model evaluation. It is noteworthy to mention that this criterion is computationally expensive when used with a larger dataset since it requires repeating the MCMC every time a data point is introduced. Based on our experiences here, other criteria such as Pareto smoothed importance sampling LFO-CV (PSIS-LFO-CV) \cite{burkner2020approximate} or widely applicable information criterion (WAIC) \cite{watanabe2010asymptotic} may be more relevant methods for large datasets. 

We have successfully demonstrated applying advanced Bayesian methods in  Davoudabadi et al. \cite{DAVOUDABADI} to more complex SOC models. We have shown the importance of these methods in inference on soil carbon dynamics, especially in scenarios where uncertainty quantification plays a significant role in carbon sequestration accounting.

In this study, we consider the effect of the microbial biomass pool on the carbon emission decomposition rate with the limitation on the maximum size of microbes, which is $5\% $ of the total SOC. Through this limitation, we have prevented too much carbon from entering the microbial pool and where excess, the extra amount is rejected by rejecting BIO state trajectories in the MCMC algorithm. Furthermore,  the precision of the single-pool statistical model of Clifford et al. \cite{clifford} has been improved upon by adding a microbial biomass and inert soil carbon pools to that model. It is possible that we could improve the growth of the population of microbes by considering a dynamic process in future studies. We could fit a model (e.g. perhaps a logistic population model with a carrying capacity) to the growth of the size of microbes. In this case, the extra amount of carbon in the BIO pool could be diverted into the other pools into which carbon could be cycled. This will be considered in future research.

\section*{Acknowledgments}
We would like to thank CSIRO for providing the data used in this study. MJD was supported by QUT-CSIRO Digital Agriculture Scholarship and a CSIRO Digital Agriculture Top-Up Scholarship. CD was supported by an Australian Research Council Discovery Project (DP200102101). We
gratefully acknowledge the computational resources provided by QUT's High Performance
Computing (HPC) and Research Support Group.

\section*{Code and Data Availability}
Dataset can be accessed online at:\\
https://doi.org/10.4225/08/54F0786D6D923.

\noindent Code for our methods and models is available at:\\ https://github.com/MJDavoudabadi/Modelling-soil-carbon-Tarlee-and-Brigalow. 



\newpage
\clearpage

\bibliography{references}

\section*{Supplementary Material}\label{SuppMater}
\appendix


\section{Notation}
The notations related to latent variables $\mathbf{X}$ at time $t$ and field $i$, their corresponding measured values $\mathbf{Y}$ and some model parameters are presented in Table \ref{TabelNotations}.

\begin{table}[!h]
\begin{center}
\small\addtolength{\tabcolsep}{-3pt}
\begin{tabular}{|c | c|} 
\hline
 \cellcolor{gray!60} Notation &  \cellcolor{gray!60} Description \\
\hline
$X_{C(t)}^i$ & The mass of SOC (t/ha)\\
\hline
$X_{W(t)}^i$ & The mass of total wheat dry
matter (t/ha)\\
\hline
$X_{S(t)}^i$ & The mass of total sorghum dry
matter (t/ha)\\
\hline
$X_{G_W(t)}^i$ & The mass of total grain dry matter produced from wheat (t/ha)\\
\hline
$X_{G_S(t)}^i$ & The mass of total grain dry matter produced from sorghum (t/ha)\\
\hline
$X_{P(t)}^i$ & The mass of total pasture dry matter (t/ha)\\
\hline
$X_{IOM(t)}^i$ & The mass of IOM (t/ha)\\
\hline
$X_{B(t)}^i$ & The mass of BIO (t/ha)\\
\hline
$X_{D(t)}^i$ & The mass of DPM (t/ha)\\
\hline
$X_{R(t)}^i$ & The mass of resistant plant material (RPM) (t/ha)\\
\hline
$X_{H(t)}^i$ & The mass of HUM (t/ha)\\
\hline
$Y_{TOC(t)}^i$ & The measured value of total SOC (t/ha)\\
\hline
$Y_{W(t)}^i$ & The measured value of total wheat dry matter (t/ha)\\
\hline
$Y_{S(t)}^i$ & The measured value of total sorghum dry matter (t/ha)\\
\hline
$Y_{G_W(t)}^i$ & The measured value of total wheat grain dry matter (t/ha)\\
\hline
$Y_{G_S(t)}^i$ & The measured value of total sorghum grain dry matter (t/ha)\\
\hline
$Y_{P(t)}^i$ & The measured value of total pasture dry matter (t/ha)\\
\hline
$Y_{IOM(t)}^i$ & The measured value of IOM (t/ha)\\
\hline
$Y_{H(t)}^i$ & The measured value of HUM (t/ha)\\
\hline
$Y_{POC(t)}^i$ & The measured value of POC (t/ha)\\
\hline
$K_C$ & The decay rate of total SOC ($Y^{-1}$)\\
\hline
$K_A$ & The decay rate of the carbon in pool $A$ ($Y^{-1}$)\\
\hline
$\pi_{AB}$ &  Proportion of the mass of carbon transfer \\ & from carbon pool $A$ to carbon pool $B$ \\
\hline
$\Delta t$ & The yearly time step \\
\hline
$P_D$ & Proportion of the carbon input that added to the DPM pool \\
\hline
\end{tabular}
\end{center}
\caption{The notations of latent variables, their corresponding measured values and some model parameters.}
\label{TabelNotations}
\end{table}

All processes and all observations at time $t$ in all fields (soil types) are denoted by $X_{(t)} = (X_{(t)}^1, X_{(t)}^2, X_{(t)}^3)$ and $Y_{(t)} = (Y_{(t)}^1, Y_{(t)}^2, Y_{(t)}^3)$, respectively. All processes at all fields (soil types) and all times are represented by $\mathbf{X}$, and $\mathbf{Y}$ represents all available data. We denote a set of variables as $Y_{1:t} = (Y_{(1)},...,Y_{(t)})$. The log-normal distribution is denoted by $LN(\mu_1,\sigma_1^2)$ with mean parameter $\mu_1$ and variance parameter $\sigma_1^2$ for a log transformation of the random variable. $N(\mu_2,\sigma_2^2)$ represents the normal distribution with mean and variance $\mu_2$ and $\sigma_2^2$, respectively.  It is important to note that in these expressions, the subscripts `1' and `2' in $\sigma_1^2$ and $\sigma_2^2$ are placeholders. In practical applications, they will be substituted with abbreviations that correspond to specific pools or models being referred to, for clearer identification and differentiation. It is the same for subscripts `A' and `B' in $K_A$ and $\pi_{AB}$ in Table \ref{TabelNotations}. Some other notations are presented wherever they are required.

\section{Prior and Proposal Distributions}\label{PriorAndProposalDists}
The model parameters and their prior probability density functions related to the Tarlee and Brigalow datasets are listed in Tables \ref{TabelPrior} and \ref{BrigalowTablePrior}. To avoid repetition, the priors of the model parameters which have the same distribution in both datasets are presented in Table \ref{TabelPrior}. Given the variety of models and their respective submodels presented in this study, we advise readers to refer to Sections \ref{SupplOneProcessModel} through \ref{SupplFiveProcessModel}. These sections provide detailed information on which parameters are associated with each specific model and submodel.

\begin{table}[!ht]
\begin{center}
\small\addtolength{\tabcolsep}{-3pt}
\begin{tabular}{|c c c|} 
\hline
 \cellcolor{gray!60} Parameter &  \cellcolor{gray!60} Prior &  \cellcolor{gray!60} Type \\
\hline
$X_{C(1978)}^1$ & Truncated-normal$(40,10^2,lower=0)$ & Uninformative\\
\hline
$X_{C(1978)}^2$ & Truncated-normal$(40,10^2,lower=0)$ & Uninformative\\
\hline
$X_{C(1978)}^3$ & Truncated-normal$(40,10^2,lower=0)$ & Uninformative\\
\hline
$X_{IOM}$ & Truncated-normal$(4,0.5^2,lower=0)$ & Uninformative\\
\hline
$K_C$ & LN$(-2.71,(0.127)^2)$ & Informative \\
\hline
$K_D$ & LN$(-2.71,(0.127)^2)$ & Informative
\\
\hline
$K_B$ &Uniform$(0,1)$ & Uninformative 
\\
\hline
$K_R$ &LN$(-2.5,(0.135)^2)$ & Informative
\\
\hline
$K_H$ &Truncated-normal$(0.02,0.01^2,lower=0)$ & Informative
\\
\hline
$c$ & N$(0.45,(0.01)^2)$ & Informative \\
\hline
$r_W$ & N$(0.5,(0.067)^2)$ & Informative \\
\hline
$r_P$ & N$(1,(0.125)^2)$ & Informative \\
\hline
$p$ & Beta$(89.9,809.1)$ & Informative \\
\hline
$h_W$ & LN$(0.825,(0.36)^2)$ & Weakly Informative \\
\hline
$\mu _{G_W}$ & N$(0.42,(1.18)^2)$ & Weakly Informative \\
\hline
$\mu _P$ & N$(1.41,(1.81)^2)$ & Weakly Informative \\
\hline
$\rho _{G_W}$ & Uniform$(-1,1)$ & Uninformative \\
\hline
$\rho _P$ & Uniform$(-1,1)$ & Uninformative \\
\hline
$\sigma _{\eta}^2$ & Inv-gamma$(0.001,0.001)$ & Uninformative \\
\hline
$\sigma _{\eta C}^2$ & Inv-gamma$(0.001,0.001)$ & Uninformative \\
\hline
$\sigma _{\eta D}^2$ & Inv-gamma$(0.001,0.001)$ & Uninformative \\
\hline
$\sigma _{\eta B}^2$ & Inv-gamma$(0.001,0.001)$ & Uninformative \\
\hline
$\sigma _{\eta R}^2$ & Inv-gamma$(0.01,0.01)$ & Uninformative \\
\hline
$\sigma _{\eta H}^2$ & Inv-gamma$(0.001,0.001)$ & Uninformative \\
\hline
$\sigma _{G_W}^2$ & Inv-gamma$(0.001,0.001)$ & Uninformative \\
\hline
$\sigma _{W}^2$ & Inv-gamma$(0.001,0.001)$ & Uninformative \\
\hline
$\sigma _{P}^2$ & Inv-gamma$(0.001,0.001)$ & Uninformative \\
\hline
$\pi _{DH}$ & Uniform$(0,1)$ & Uninformative \\
\hline
$\pi _{RH}$ & Uniform$(0,1)$ & Uninformative \\
\hline
$\pi _{HH}$ & Uniform$(0,1)$ & Uninformative \\
\hline
$\pi _{BH}$ & Uniform$(0,1)$ & Uninformative \\
\hline
$\pi _{DB}$ & Uniform$(0,1)$ & Uninformative \\
\hline
$\pi _{RB}$ & Uniform$(0,1)$ & Uninformative \\
\hline
$\pi _{HB}$ & Uniform$(0,1)$ & Uninformative \\
\hline
$\pi _{BB}$ & Uniform$(0,1)$ & Uninformative \\
\hline
$\pi _{CB}$ & Uniform$(0,1)$ & Uninformative \\
\hline
$\pi _{BC}$ & Uniform$(0,1)$ & Uninformative \\
\hline
$\sigma _{\epsilon TOC}^2$ & 0.025 & Fixed \\
\hline
$\sigma _{\epsilon POC}^2$ & 0.9 & Fixed \\
\hline
$\sigma _{\epsilon G_W}^2$ & 0.023 & Fixed \\
\hline
$\sigma _{\epsilon W}^2$ & 0.133 & Fixed \\
\hline
$\sigma _{\epsilon P}^2$ & 0.067 & Fixed \\
\hline
$\sigma _{\epsilon IOM}^2$ & 0.01 & Fixed \\
\hline
$\sigma _{\epsilon H}^2$ & 0.1 & Fixed \\
\hline
\end{tabular}
\end{center}
\caption{Prior distributions of parameters of the Tarlee dataset and the ones are common in both datasets.}
\label{TabelPrior}
\end{table}

\begin{table}[!h]
\begin{center}
\small\addtolength{\tabcolsep}{-3pt}
\begin{tabular}{|c c c|} 
\hline
 \cellcolor{gray!60} Parameter &  \cellcolor{gray!60} Prior &  \cellcolor{gray!60} Type \\
\hline
$X_{C(1981)}^1$ & Truncated-normal$(60,15^2,lower=0)$ & Uninformative\\
\hline
$X_{C(1981)}^2$ & Truncated-normal$(60,15^2,lower=0)$ & Uninformative\\
\hline
$X_{C(1981)}^3$ & Truncated-normal$(60,15^2,lower=0)$ & Uninformative\\
\hline
$X_{IOM}$ & Truncated-normal$(12,2^2,lower=0)$ & Uninformative\\
\hline
$h_S$ & LN$(0.46, (1.6)^2)$ & Informative \\
\hline
$\rho _{G_S}$ & Uniform$(-1,1)$ & Uninformative \\
\hline
$\mu _{G_S}$ & N$(0.42,(1.18)^2)$ & Weakly Informative \\
\hline
$r_S$ & N$(0.5,(0.067)^2)$ & Informative \\
\hline
$K_B$ &Truncated-normal$(0.66,0.3^2,lower=0)$ & Informative 
\\
\hline
$K_R$ &Truncated-normal$(0.15,0.075^2,lower=0)$ & Informative \\
\hline
$\sigma _{G_S}^2$ & Inv-gamma$(0.001,0.001)$ & Uninformative \\
\hline
$\sigma _{\eta B}^2$ & Truncated-normal$(0,0.5^2,lower=0)$ & Weakly Informative\\
\hline
$\sigma _{\eta R}^2$ & Truncated-normal$(0,0.5^2,lower=0)$ & Weakly Informative\\
\hline
$\sigma _{\eta H}^2$ & Truncated-normal$(0,0.5^2,lower=0)$ & Weakly Informative\\
\hline
$\sigma _{S}^2$ & Inv-gamma$(0.001,0.001)$ & Uninformative \\
\hline
$\sigma _{\epsilon G_S}^2$ & 0.023 & Fixed \\
\hline
$\sigma _{\epsilon S}^2$ & 0.133 & Fixed \\
\hline
\end{tabular}
\end{center}
\caption{Prior distributions of parameters of the Brigalow dataset.}
\label{BrigalowTablePrior}
\end{table}

The proposal density functions of the one, two, and three-pool models applied on the Tarlee dataset are presented in Table \ref{TabPropDist123}. In the three-pool model, the proposal density functions of some parameters are different from the ones in the one and two-pool models, therefore, we show them in Table \ref{TabPropDist3}. The proposal density functions of the five-pool model related to the Tarlee dataset are shown in Table \ref{TabelProposal}. Tables \ref{Brig1PoolProp}, \ref{Brig2PoolProp}, \ref{Brig3PoolProp}, and \ref{Brig5PoolProp} show respectively the proposal density functions of the one, two, three, and five-pool models related to the Brigalow dataset. To avoid repetition, the proposal density functions of parameters which are the same in two or three-pool model are not shown in Tables \ref{Brig3PoolProp} and \ref{Brig5PoolProp}.

\begin{table}[!h]
\begin{center}
\small\addtolength{\tabcolsep}{-3pt}
\begin{tabular}{|c c|} 
\hline
 \cellcolor{gray!60} Parameter &  \cellcolor{gray!60} Proposal\\
\hline
$K_C$ & N$(K_C, 0.001^2)$\\
\hline
$K_B$ & Truncated-normal$(K_B, 0.09^2, lower = 0)$\\
\hline
$c$ & Truncated-normal$(c, 0.005^2, lower = 0, upper = 1)$\\
\hline
$r_W$ & Truncated-normal$(r_W, 0.05^2, lower = 0)$\\
\hline
$r_P$ & Truncated-normal$(r_P, 0.05^2, lower = 0)$\\
\hline
$p$ & Truncated-normal$(p, 0.005^2, lower = 0, upper = 1)$\\
\hline
$h_W$ & Truncated-normal$(h_W, 0.05^2, lower = 0)$\\
\hline
$\mu_{G_W}$ & N$(\mu_{G_W}, 0.05^2)$\\
\hline
$\mu_P$ & N$(\mu_P, 0.05^2)$\\
\hline
$\rho_{G_W}$ & Truncated-normal$(\rho_{G_W}, 0.05^2, lower = -1, upper = 1)$\\
\hline
$\rho_P$ & Truncated-normal$(\rho_P, 0.1^2, lower = -1, upper = 1)$\\
\hline
$\sigma_{\eta}^2$ & Truncated-normal$(\sigma_{\eta}^2, 0.001^2, lower = 0)$\\
\hline
$\sigma_{\eta C}^2$ & Truncated-normal$(\sigma_{\eta C}^2, 0.001^2, lower = 0)$\\
\hline
$\sigma_{\eta B}^2$ & Truncated-normal$(\sigma_{\eta B}^2, 0.01^2, lower = 0)$\\
\hline
$\sigma_{G_W}^2$ & Truncated-normal$(\sigma_{G_W}^2, \frac{\sigma_{G_W}^2}{20^2}, lower = 0)$\\
\hline
$\sigma_{W}^2$ & Truncated-normal$(\sigma_{W}^2, 0.001^2, lower = 0)$\\
\hline
$\sigma_{P}^2$ & Truncated-normal$(\sigma_{P}^2, 0.1^2, lower = 0)$\\
\hline
$\pi_{CB}$ & Truncated-normal$(\pi_{CB}, 0.05^2, lower = 0, upper = 1)$\\
\hline
$\pi_{BC}$ & Truncated-normal$(\pi_{BC}, 0.1^2, lower = 0, upper = 1)$\\
\hline
$\pi_{BB}$ & Truncated-normal$(\pi_{BB}, 0.1^2, lower = 0, upper = 1)$\\
\hline
$X_{C(1978)}^1$ & Truncated-normal$(X_{C(1978)}^1, 5^2, lower = 0)$\\
\hline
$X_{C(1978)}^2$ & Truncated-normal$(X_{C(1978)}^2, 5^2, lower = 0)$\\
\hline
$X_{C(1978)}^3$ & Truncated-normal$(X_{C(1978)}^3, 5^2, lower = 0)$\\
\hline
$X_{IOM}$ & Truncated-normal$(X_{IOM}, 0.9^2, lower = 0)$\\
\hline
\end{tabular}
\end{center}
\caption{Proposal distributions of one, two, and three-pool models used in the CPM method for the Tarlee dataset.}
\label{TabPropDist123}
\end{table}

\begin{table}[!h]
\begin{center}
\small\addtolength{\tabcolsep}{-3pt}
\begin{tabular}{|c c|} 
\hline
 \cellcolor{gray!60} Parameter &  \cellcolor{gray!60} Proposal\\
\hline
$K_C$ & N$(K_C, 0.005^2)$\\
\hline
$\mu_G$ & N$(\mu_G, 0.075^2)$\\
\hline
$\mu_P$ & N$(\mu_P, 0.1^2)$\\
\hline
$X_{C(1978)}^1$ & Truncated-normal$(X_{C(1978)}^1, 2^2, lower = 0)$\\
\hline
$X_{C(1978)}^2$ & Truncated-normal$(X_{C(1978)}^2, 2^2, lower = 0)$\\
\hline
$X_{C(1978)}^3$ & Truncated-normal$(X_{C(1978)}^3, 2^2, lower = 0)$\\
\hline
$X_{IOM}$ & Truncated-normal$(X_{IOM}, 0.09^2, lower = 0)$\\
\hline
\end{tabular}
\end{center}
\caption{Proposal distributions of some parameters in the three-pool model used in the CPM method for the Tarlee dataset.}
\label{TabPropDist3}
\end{table}

\begin{table}[!h]
\begin{center}
\small\addtolength{\tabcolsep}{-3pt}
\begin{tabular}{|c c|} 
\hline
 \cellcolor{gray!60} Parameter &  \cellcolor{gray!60} Proposal\\
\hline
$K_D$ & Truncated-normal$(K_D, 0.005^2, lower = 0)$\\
\hline
$K_B$ & Truncated-normal$(K_B, 0.09^2, lower = 0)$\\
\hline
$K_R$ & Truncated-normal$(K_R, 0.005^2, lower = 0)$\\
\hline
$K_H$ & Truncated-normal$(K_H, 0.006^2, lower = 0)$\\
\hline
c & Truncated-normal$(c, 0.005^2, lower = 0, upper = 1)$\\
\hline
$r_W$ & Truncated-normal$(r_W, 0.05^2, lower = 0)$\\
\hline
$r_P$ & Truncated-normal$(r_P, 0.05^2, lower = 0)$\\
\hline
p & Truncated-normal$(p, 0.005^2, lower = 0, upper = 1)$\\
\hline
h & Truncated-normal$(h, 0.05^2, lower = 0)$\\
\hline
$\mu_G$ & N$(K, 0.075^2)$\\
\hline
$\mu_P$ & N$(K, 0.1^2)$\\
\hline
$\rho_G$ & Truncated-normal$(\rho_G, 0.25^2, lower = -1, upper = 1)$\\
\hline
$\rho_P$ & Truncated-normal$(\rho_P, 0.2^2, lower = -1, upper = 1)$\\
\hline
$\sigma_{\eta D}^2$ & Truncated-normal$(\sigma_{\eta D}^2, 0.1^2, lower = 0)$\\
\hline
$\sigma_{G}^2$ & Truncated-normal$(\sigma_{G}^2, \frac{\sigma_{G}^2}{20^2}, lower = 0)$\\
\hline
$\sigma_{W}^2$ & Truncated-normal$(\sigma_{W}^2, 0.01^2, lower = 0)$\\
\hline
$\sigma_{P}^2$ & Truncated-normal$(\sigma_{P}^2, 0.1^2, lower = 0)$\\
\hline
$X_{IOM}$ & Truncated-normal$(X_{IOM}, 0.09^2, lower = 0)$\\
\hline
$X_{C(1978)}^1$ & Truncated-normal$(X_{C(1978)}^1, 1.3^2, lower = 0)$\\
\hline
$X_{C(1978)}^2$ & Truncated-normal$(X_{C(1978)}^2, 1.3^2, lower = 0)$\\
\hline
$X_{C(1978)}^3$ & Truncated-normal$(X_{C(1978)}^3, 1.3^2, lower = 0)$\\
\hline
$\sigma_{\eta B}^2$ & Truncated-normal$(\sigma_{B}^2, 0.1^2, lower = 0)$\\
\hline
$P_{D}$ & Truncated-normal$(P_{D}, 0.15^2, lower = 0, upper = 1)$\\
\hline
$\sigma_{R}^2$ & Truncated-normal$(\sigma_{R}^2, 0.09^2, lower = 0)$\\
\hline
$\sigma_{H}^2$ & Truncated-normal$(\sigma_{H}^2, 0.01^2, lower = 0)$\\
\hline
$\pi_{DB}$ & Truncated-normal$(\pi_{DB}, 0.05^2, lower = 0, upper = 1)$\\
\hline
$\pi_{BB}$ & Truncated-normal$(\pi_{BB}, 0.15^2, lower = 0, upper = 1)$\\
\hline
$\pi_{DH}$ & Truncated-normal$(\pi_{DH}, 0.09^2, lower = 0, upper = 1)$\\
\hline
$\pi_{RH}$ & Truncated-normal$(\pi_{RH}, 0.1^2, lower = 0, upper = 1)$\\
\hline
$\pi_{HH}$ & Truncated-normal$(\pi_{HH}, 0.1^2, lower = 0, upper = 1)$\\
\hline
$\pi_{BH}$ & Truncated-normal$(\pi_{BH}, 0.015^2, lower = 0, upper = 1)$\\
\hline
$\pi_{RB}$ & Truncated-normal$(\pi_{RB}, 0.09^2, lower = 0, upper = 1)$\\
\hline
$\pi_{HB}$ & Truncated-normal$(\pi_{HB}, 0.1^2, lower = 0, upper = 1)$\\
\hline
\end{tabular}
\end{center}
\caption{Proposal distributions of the five-pool model used in the CPM method for the Tarlee dataset.}
\label{TabelProposal}
\end{table}

\begin{table}[!h]
\begin{center}
\small\addtolength{\tabcolsep}{-3pt}
\begin{tabular}{|c c|} 
\hline
 \cellcolor{gray!60} Parameter &  \cellcolor{gray!60} Proposal\\
\hline
$K_C$ & N$(K_C, 0.002^2)$\\
\hline
$c$ & Truncated-normal$(c, 0.005^2, lower = 0, upper = 1)$\\
\hline
$r_W$ & Truncated-normal$(r_W, 0.05^2, lower = 0)$\\
\hline
$p$ & Truncated-normal$(p, 0.005^2, lower = 0, upper = 1)$\\
\hline
$h_W$ & Truncated-normal$(h_W, 0.05^2, lower = 0)$\\
\hline
$\mu_{G_W}$ & N$(\mu_{G_W}, 0.05^2)$\\
\hline
$\rho_{G_W}$ & Truncated-normal$(\rho_{G_W}, 0.05^2, lower = -1, upper = 1)$\\
\hline
$\sigma_{\eta}^2$ & Truncated-normal$(\sigma_{\eta}^2, 0.001^2, lower = 0)$\\
\hline
$\sigma_{G_W}^2$ & Truncated-normal$(\sigma_{G_W}^2, \frac{\sigma_{G_W}^2}{20^2}, lower = 0)$\\
\hline
$\sigma_{W}^2$ & Truncated-normal$(\sigma_{W}^2, 0.001^2, lower = 0)$\\
\hline
$X_{C(1981)}^1$ & Truncated-normal$(X_{C(1981)}^1, 2^2, lower = 0)$\\
\hline
$X_{C(1981)}^2$ & Truncated-normal$(X_{C(1981)}^2, 2^2, lower = 0)$\\
\hline
$X_{C(1981)}^3$ & Truncated-normal$(X_{C(1981)}^3, 2^2, lower = 0)$\\
\hline
$r_S$ & Truncated-normal$(r_S, 0.05^2, lower = 0)$\\
\hline
$\sigma_{S}^2$ & Truncated-normal$(\sigma_{S}^2, 0.05^2, lower = 0)$\\
\hline
$\mu_{G_S}$ & N$(\mu_{G_S}, 0.05^2)$\\
\hline
$\rho_{G_S}$ & Truncated-normal$(\rho_{G_S}, 0.25^2, lower = -1, upper = 1)$\\
\hline
$\sigma_{G_S}^2$ & Truncated-normal$(\sigma_{G_S}^2, \frac{\sigma_{G_S}^2}{20^2}, lower = 0)$\\
\hline
$h_S$ & Truncated-normal$(h_S, 0.25^2, lower = 0)$\\
\hline
\end{tabular}
\end{center}
\caption{Proposal distributions of one-pool models used in the CPM method for the Brigalow dataset.}
\label{Brig1PoolProp}
\end{table}

\begin{table}[!h]
\begin{center}
\small\addtolength{\tabcolsep}{-3pt}
\begin{tabular}{|c c|} 
\hline
 \cellcolor{gray!60} Parameter &  \cellcolor{gray!60} Proposal\\
\hline
$K_C$ & N$(K_C, 0.01^2)$\\
\hline
$c$ & Truncated-normal$(c, 0.005^2, lower = 0, upper = 1)$\\
\hline
$r_W$ & Truncated-normal$(r_W, 0.05^2, lower = 0)$\\
\hline
$p$ & Truncated-normal$(p, 0.005^2, lower = 0, upper = 1)$\\
\hline
$h_W$ & Truncated-normal$(h_W, 0.1^2, lower = 0)$\\
\hline
$\mu_{G_W}$ & N$(\mu_{G_W}, 0.05^2)$\\
\hline
$\rho_{G_W}$ & Truncated-normal$(\rho_{G_W}, 0.05^2, lower = -1, upper = 1)$\\
\hline
$\sigma_{\eta}^2$ & Truncated-normal$(\sigma_{\eta}^2, 0.001^2, lower = 0)$\\
\hline
$\sigma_{G_W}^2$ & Truncated-normal$(\sigma_{G_W}^2, \frac{\sigma_{G_W}^2}{20^2}, lower = 0)$\\
\hline
$\sigma_{W}^2$ & Truncated-normal$(\sigma_{W}^2, 0.01^2, lower = 0)$\\
\hline
$X_{C(1981)}^1$ & Truncated-normal$(X_{C(1981)}^1, 2^2, lower = 0)$\\
\hline
$X_{C(1981)}^2$ & Truncated-normal$(X_{C(1981)}^2, 2^2, lower = 0)$\\
\hline
$X_{C(1981)}^3$ & Truncated-normal$(X_{C(1981)}^3, 2^2, lower = 0)$\\
\hline
$X_{IOM}$ & Truncated-normal$(X_{IOM}, 0.05^2, lower = 0)$\\
\hline
$r_S$ & Truncated-normal$(r_S, 0.05^2, lower = 0)$\\
\hline
$\sigma_{S}^2$ & Truncated-normal$(\sigma_{S}^2, 0.01^2, lower = 0)$\\
\hline
$\mu_{G_S}$ & N$(\mu_{G_S}, 0.05^2)$\\
\hline
$\rho_{G_S}$ & Truncated-normal$(\rho_{G_S}, 0.25^2, lower = -1, upper = 1)$\\
\hline
$\sigma_{G_S}^2$ & Truncated-normal$(\sigma_{G_S}^2, \frac{\sigma_{G_S}^2}{20^2}, lower = 0)$\\
\hline
$h_S$ & Truncated-normal$(h_S, 0.25^2, lower = 0)$\\
\hline
\end{tabular}
\end{center}
\caption{Proposal distributions of two-pool models used in the CPM method for the Brigalow dataset.}
\label{Brig2PoolProp}
\end{table}

\begin{table}[!h]
\begin{center}
\small\addtolength{\tabcolsep}{-3pt}
\begin{tabular}{|c c|} 
\hline
 \cellcolor{gray!60} Parameter &  \cellcolor{gray!60} Proposal\\
\hline
$\sigma_{\eta_C}^2$ & Truncated-normal$(\sigma_{\eta_C}^2, 0.001^2, lower = 0)$\\
\hline
$\sigma_{B}^2$ & Truncated-normal$(\sigma_{B}^2, 0.1^2, lower = 0)$\\
\hline
$K_B$ & Truncated-normal$(K_B, 0.09^2, lower = 0)$\\
\hline
$\pi_{DB}$ & Truncated-normal$(\pi_{DB}, 0.1^2, lower = 0)$\\
\hline
$\pi_{BB}$ & Truncated-normal$(\pi_{BB}, 0.1^2, lower = 0)$\\
\hline
$\pi_{BC}$ & Truncated-normal$(\pi_{BC}, 0.1^2, lower = 0)$\\
\hline
$\pi_{CB}$ & Truncated-normal$(\pi_{CB}, 0.05^2, lower = 0)$\\
\hline
$K_{R}$ & Truncated-normal$(K_{R}, 0.1^2, lower = 0)$\\
\hline
\end{tabular}
\end{center}
\caption{Proposal distributions of three-pool models used in the CPM method for the Brigalow dataset.}
\label{Brig3PoolProp}
\end{table}

\begin{table}[!h]
\begin{center}
\small\addtolength{\tabcolsep}{-3pt}
\begin{tabular}{|c c|} 
\hline
 \cellcolor{gray!60} Parameter &  \cellcolor{gray!60} Proposal\\
\hline
$\sigma_{\eta_D}^2$ & Truncated-normal$(\sigma_{\eta_D}^2, 0.02^2, lower = 0)$\\
\hline
$\sigma_{\eta_B}^2$ & Truncated-normal$(\sigma_{\eta_B}^2, 0.1^2, lower = 0)$\\
\hline
$\sigma_{\eta_R}^2$ & Truncated-normal$(\sigma_{\eta_R}^2, 0.9^2, lower = 0)$\\
\hline
$\sigma_{\eta_H}^2$ & Truncated-normal$(\sigma_{\eta_H}^2, 0.1^2, lower = 0)$\\
\hline
$\sigma_{G_W}^2$ & Truncated-normal$(\sigma_{G_W}^2, \frac{\sigma_{G_W}^2}{30^2}, lower = 0)$\\
\hline
$\sigma_{G_S}^2$ & Truncated-normal$(\sigma_{G_S}^2, \frac{\sigma_{G_W}^2}{10^2}, lower = 0)$\\
\hline
$\sigma_{S}^2$ & Truncated-normal$(\sigma_{S}^2, 0.05^2, lower = 0)$\\
\hline
$X_{IOM}$ & Truncated-normal$(X_{IOM}, 0.9^2, lower = 0)$\\
\hline
$X_{C(1981)}^1$ & Truncated-normal$(X_{C(1981)}^1, 3^2, lower = 0)$\\
\hline
$X_{C(1981)}^2$ & Truncated-normal$(X_{C(1981)}^2, 3^2, lower = 0)$\\
\hline
$X_{C(1981)}^3$ & Truncated-normal$(X_{C(1981)}^3, 3^2, lower = 0)$\\
\hline
$K_D$ & Truncated-normal$(K_D, 0.005^2, lower = 0)$\\
\hline
$K_B$ & Truncated-normal$(K_B, 0.1^2, lower = 0)$\\
\hline
$K_R$ & Truncated-normal$(K_R, 0.05^2, lower = 0)$\\
\hline
$K_H$ & Truncated-normal$(K_H, 0.01^2, lower = 0)$\\
\hline
$h_W$ & Truncated-normal$(h_W, 0.05^2, lower = 0)$\\
\hline
$h_S$ & Truncated-normal$(h_S, 0.5^2, lower = 0)$\\
\hline
$P_D$ & Truncated-normal$(P_D, 0.15^2, lower = 0)$\\
\hline
$\pi_{DH}$ & Truncated-normal$(\pi_{DH}, 0.09^2, lower = 0)$\\
\hline
$\pi_{RH}$ & Truncated-normal$(\pi_{RH}, 0.1^2, lower = 0)$\\
\hline
$\pi_{HH}$ & Truncated-normal$(\pi_{HH}, 0.1^2, lower = 0)$\\
\hline
$\pi_{BH}$ & Truncated-normal$(\pi_{BH}, 0.15^2, lower = 0)$\\
\hline
$\pi_{RB}$ & Truncated-normal$(\pi_{RB}, 0.09^2, lower = 0)$\\
\hline
$\pi_{HB}$ & Truncated-normal$(\pi_{HB}, 0.1^2, lower = 0)$\\
\hline
$\rho_{G_S}$ & Truncated-normal$(\rho_{G_S}, 0.1^2, lower = -1, upper = 1)$\\
\hline
\end{tabular}
\end{center}
\caption{Proposal distributions of five-pool models used in the CPM method for the Brigalow dataset.}
\label{Brig5PoolProp}
\end{table}

\section{State-space Model}\label{SSM}
The state-space model uses observable measurement variable $Y_{(t)}$ and unobserved state variable $X_{(t)}$ which can be estimated through observational data that depend on the state variable, to describe a system. This model includes the first two levels of the hierarchy of the BHM framework and its generic representation with Gaussian noise is 
\begin{equation} \label{eq1SS}
\begin{split}
&X_{(t)}=f(X_{(t-1)})+ \boldsymbol{B} u_{(t)} + \epsilon_{(t)} \\
&Y_{(t)}=g(X_{(t)})+ \nu_{(t)}  ;
\end{split}
\end{equation}
where $\epsilon_{(t)} \sim N(\mu , \sigma _X^2)$ and $\nu_{(t)} \sim N(\lambda , \sigma _Y^2)$ are state and measurement noise components, respectively and the control-input matrix $\boldsymbol{B}$ is applied to a known vector of inputs $u_{(t)}$. The aim is to produce estimators for the state variable $X_{(t)}$ through the filtering distribution $p(X_{(t)} \vert Y_{1:t})$. When the filtering distribution does not have a closed-form expression some methods such as the Kalman filter (KF) (in the case of linear-Gaussian model) and particle filter can be used to approximate it. For the sake of simplicity, we assume the static parameter $\boldsymbol{\theta}$ is fixed in this section.

\subsection{Kalman Filter} \label{Sub.KF}

In the case of linear-Gaussian state-space model, the KF can be used to estimate state variable as an efficient method since it is an  optimal estimator in the sense of minimising the variance of the estimated state. The linear-Gaussian state-space model has the form 
\begin{align*}
    &X_{(t)}= \boldsymbol{A}^* X_{(t-1)}+ \boldsymbol{B}^* u_{(t)} + \epsilon_{(t)}^* \\
    &Y_{(t)}= \boldsymbol{C}^* X_{(t)} + \nu_{(t)}^* ;
\end{align*}
where $\epsilon_{(t)}^* \sim N(\boldsymbol{0},\boldsymbol{Q}^*)$, $\nu_{(t)}^* \sim N(\boldsymbol{0},\boldsymbol{R}^*)$, $\boldsymbol{A}^*$ is the state-transition matrix, the control-input matrix $\boldsymbol{B}^*$ is applied to a known vector of inputs $u_{(t)}$, and $\boldsymbol{C}^*$  is the observation matrix. The KF method is shown in Algorithm \ref{euclidKF1}, here $\mbox{MVN}(\boldsymbol{\mu}, \boldsymbol{\Sigma})$ denotes the multivariate normal density with mean vector $\boldsymbol{\mu}$ and covariance matrix $\boldsymbol{\Sigma}$. In Algorithm \ref{euclidKF1}, $\boldsymbol{K}_{(t)}$ is the Kalman gain matrix, $\boldsymbol{P}_{(t)}^{t-1}$ and $X_{(t)}^{t-1}$ are the process noise and the expectations of state variable, respectively given all observations up to and including time $t-1$.

\begin{algorithm}
\caption{Kalman filter algorithm}\label{euclidKF1}
\begin{algorithmic}[1]

\State Initialize with initial state $\Hat{X}_{(0)} = x_{(0)}$ and $\Hat{\boldsymbol{P}}_{(0)} = \boldsymbol{Q}^*$ at $t=0$;
\For {$t = 1,...,\textit{T}$}
\State $X_{(t)}^{t-1} = \boldsymbol{A}^* \Hat{X}_{(t-1)} + \boldsymbol{B}^* u_{(t)}$,  \quad \text{State estimate extrapolation};
\State $\boldsymbol{P}_{(t)}^{t-1} = \boldsymbol{A}^* \Hat{\boldsymbol{P}}_{(t-1)}\boldsymbol{A}^{*'}  + \boldsymbol{Q}^*$,  \quad \text{State covariance extrapolation};
\State $\boldsymbol{K}_{(t)} = \boldsymbol{P}_{(t)}^{t-1} \boldsymbol{C}^{*'}[\boldsymbol{R}^* + \boldsymbol{C}^* \boldsymbol{P}_{(t)}^{t-1} \boldsymbol{C}^{*'}]^{-1}$,  \quad \text{Kalman gain matrix};
\State $\Hat{X}_{(t)} = X_{(t)}^{t-1} + \boldsymbol{K}_{(t)} [Y_{(t)} - \boldsymbol{C}^* X_{(t)}^{t-1}]$, \quad \text{State estimate update};
\State $\Hat{\boldsymbol{P}}_{(t)} = [\boldsymbol{I} - \boldsymbol{K}_{(t)} \boldsymbol{C}^*] \boldsymbol{P}_{(t)}^{t-1} $,        \quad \text{State covariance update};
\State Compute the log-likelihood contribution, $l_{(t)}^{\mathrm{KF}}$, at time $t$ through the density $\mbox{MVN}(Y_{(t)} - \boldsymbol{C}^* X_{(t)}^{t-1}, \boldsymbol{R}^* + \boldsymbol{C}^* \boldsymbol{P}_{(t)}^{t-1} \boldsymbol{C}^{*'})$;
\EndFor
\State The complete log-likelihood can be calculated as $L^* = \sum_t l_{(t)}^{\mathrm{KF}}$
\end{algorithmic}
\end{algorithm}
To apply the KF to the Tarlee model, we use  a log transformation, i.e. $X^* = \log(X)$ and $Y^* = \log(Y)$ in order to form a linear-Gaussian state-space model for the state-space model formed by the sub-model involving $\lbrace X^*_G, X^*_W, X^*_P, Y^*_G, Y^*_W, Y^*_P \rbrace$. Since $X^*_{W(t)}$ depends on $X^*_{G(t)}$ and because of the auto-regressive structure of the state-space model, we rewrite $X^*_{W(t)}$ as follows
\begin{align*}
    &X_{W(t)}^{*i} \sim N(\log h + \mu _G + \rho _G (X_{G(t-1)}^{*i}- \mu _G), \sigma_W^2 + \sigma_G^2).
\end{align*}

We can produce estimators of the state variables in the case of linear and non-linear state-space models through particle filters. The algorithm of one of particle filters is introduced in the next section.

\subsection{Bootstrap Particle Filter}\label{Sub.BPF}

In the linear and non-linear cases, particle filters can be used to produce estimators of the state variable $X_{(t)}$ and the simplest form of particle filters known as bootstrap particle filter is applied on non-linear part of the model in this study. The algorithm of bootstrap particle filter is provided in Algorithm \ref{euclidBF1}.
\begin{algorithm}
\caption{Bootstrap particle filter algorithm}\label{euclidBF1}
\begin{algorithmic}[1]

\For {$k = 1,...,\textit{N}$}
\State $t=1$, \text{draw sample} $X^{k}_{(1)} \sim p(X_{(1)})$;
\EndFor
\For {$t = 2,...,\textit{T}$}

\For {$k = 1,...,\textit{N}$}
\State Draw sample $X_{(t)}^k \sim p(X_{(t)} \vert X^{*k}_{(t-1)})$; 
\State Calculate weights $w_{(t)}^k = p(Y_{(t)} \vert X_{(t)}^k)$;
\EndFor
\State Estimate the log-likelihood component for the $t^{th}$ observation, $\hat{l}_{(t)} = \log \left(\dfrac{\sum_j w_{(t)}^j}{N}\right)$;

\State Normalise weights $W_{(t)}^k = \dfrac{w_{(t)}^k}{\sum_j w_{(t)}^j}$ for $k \in \{1, 2, \dots, N \}$;

\State Resample with replacement $N$ particles $X_{(t)}^k$ based on the normalised importance weights;
\State Estimate the overall log-likelihood  $L^* = \sum_t \hat{l}_{(t)}$.

\EndFor

\end{algorithmic}
\end{algorithm}

\subsection{Correlated Pseudo-marginal Method}\label{subsecCPM}
When the process and observation models depend on a set of unknown static parameters $\boldsymbol{\theta}$, one can treat the parameters as random variables and utilise Bayesian approach to estimate $\boldsymbol{\theta}$. In the case of having an intractable posterior distribution, numerical methods such as Markov chain Monte Carlo (MCMC) methods can be used. In this study, we utilise correlated pseudo-marginal method, one of the MCMC methods, to generate a sequence of correlated random samples from a probability distribution from which direct sampling is difficult. The likelihood estimators $\hat{p}(\mathbf{Y} \vert \mathbf{X}, \boldsymbol{\theta} ^*)$ and $\hat{p}(\mathbf{Y} \vert \mathbf{X}, \boldsymbol{\theta}_{m-1} )$ in the acceptance ratio of the CPM method are correlated through correlating the auxiliary random numbers $U$, used to obtain these estimators, in order to reduce the variance of the resulting ratio. The CPM algorithm is presented in Algorithm \ref{CPMalgorithm}. To have a highly correlated likelihood estimators in the CPM algorithm it is required to use a particle filter that  processes  the  random  numbers  in  such  a  manner  that  the  likelihood  estimates  are  similar as possible when slightly perturbing the random numbers. Algorithm \ref{BPF_CPM} shows the particle filter with a given set of random numbers.


\begin{algorithm}[ht]
\caption{Correlated pseudo-marginal algorithm}\label{CPMalgorithm}
\begin{algorithmic}[1]
\State Initialise $\boldsymbol{\theta}_0$;
\For {$m = 1,...,\textit{M}^*$}
\State Sample $\boldsymbol{\theta}^{*} \sim Q(.\vert \boldsymbol{\theta}_{m-1})$;
\State Sample $\xi \sim N(\textbf{0}, \boldsymbol{I})$ and set $U^* = \tau U_{m-1} + \sqrt{1-\tau ^2} \xi$;
\State Compute the estimator $\hat{p} (\mathbf{Y} \vert  \boldsymbol{\theta} ^*, U^*)$ using Algorithm \ref{BPF_CPM}
\State Compute the acceptance ratio:
\begin{align*}
r=\frac{\hat{p} (\mathbf{Y} \vert  \boldsymbol{\theta} ^*, U^*) p(\boldsymbol{\theta} ^*)Q(\boldsymbol{\theta}_{m-1} \vert \boldsymbol{\theta}^*)}{\hat{p}(\mathbf{Y} \vert  \boldsymbol{\theta}_{m-1}, U_{m-1} ) p(\boldsymbol{\theta}_{m-1} )Q(\boldsymbol{\theta}^*\vert \Theta_{m-1} )};
\end{align*}

\State Accept $(\boldsymbol{\theta} ^*, U^*)$ with probability $\min (r,1)$ otherwise, output $(\boldsymbol{\theta}_{m-1}, U_{m-1})$
\EndFor
\end{algorithmic}
\end{algorithm}

\begin{algorithm}
\caption{Particle filter with fixed random numbers}\label{BPF_CPM}
\begin{algorithmic}[1]

\State Sample $U_{(j^*)} \sim N(0,1)$ and $V_{(i^*)} \sim N(0,1)$ for all $j^* \in \{1, \dots, TN \}$ and $i^* \in \{ 1, \dots, T \}$;
\State Sample $X_{(1)}^k \sim p(. \vert U_{1:N}, \boldsymbol{\theta})$ for all $k \in \{1, \dots, N\}$;
\For {$t = 1,...,\textit{T-1}$}
\State Sort the collection $\{X_{(t)}^1, \dots, X_{(t)}^N\}$;
\State Compute importance weights $w_{(t)}^k$ and log-likelihoods $\hat{l}_{(t)} = \log \left(\dfrac{\sum_k w_{(t)}^k}{N}\right)$ for $k \in \{1, \dots, N\}$;
\State Sample $X_{(t)}^k$ based on systematic resampling using random values $V_{1:T}$ and normalised 
\newline \phantom{aa} weights $W_{(t)}^k$ for $k \in \{1, \dots, N\}$;
\State Set $X_{(t+1)}^k$ as a sample from $p(. \vert X_{(t)}^k, U_{Nt+1:N(t + 1)}, \boldsymbol{\theta})$ for $k \in \{1, \dots, N\}$;

\EndFor
\State Estimate the overall log-likelihood  $L^* = \sum_t \hat{l}_{(t)}$.

\end{algorithmic}
\end{algorithm}

The process and observation models of the Tarlee and Brigalow datasets are presented in the next sections.
\section{One-pool Model}\label{SupplOneProcessModel}

\subsection{Process Model}\label{OneProcessModel}
The process model of the one-pool model at time $t$ in field (or soil type) $i$ is: 
\begin{align}
     &\log (X_{C(t)}^i) = \log (X_{C(t-1)}^i e^{-K_C\Delta t} + I_{C(t)}^i) + \eta _{(t)}^i, \ \ \ \eta_{(t)}^i \sim N(0, \sigma_{\eta} ^2);  \label{eq1S}\\
    &X_{G_W(t)}^i \sim LN(\mu _{G_W} + \rho _{G_W} (\log (X_{G_W(t-1)}^i)- \mu _{G_W}), \sigma_{G_W}^2);  \label{eq2S}\\
    &X_{W(t)}^i \sim LN(\log h_W + \log (x_{G_W(t)}^i), \sigma_W^2);      \label{eq3S}\\
    &X_{P(t)}^i \sim LN(\mu _P + \rho _P (\log (X_{P(t-1)}^i)- \mu _P), \sigma_P^2); \label{eq4S}\\
    &X_{G_S(t)}^i \sim LN(\mu _{G_S} + \rho _{G_S} (\log (X_{G_S(t-1)}^i)- \mu _{G_S}), \sigma_{G_S}^2); \quad \text{and} \label{eq5sS}\\
    &X_{S(t)}^i \sim LN(\log h_S + \log (x_{G_S(t)}^i), \sigma_S^2); \label{eq6sS} 
\end{align}
\noindent where $h_W$, $h_S$, $\rho _P$, $\rho _{G_W}$, and $\rho _{G_S}$ denote the harvest index which is the ratio of wheat to grain, the harvest index which is the ratio of sorghum to grain, auto-regressive parameters for the evolution of pasture total dry matter (TDM) and grain TDM of wheat and sorghum, respectively. Notice that $X_{W(t)}^i$ and $X_{S(t)}^i$ are defined conditional on $X_{G_W(t)}^i = x_{G_W(t)}^i$ and $X_{G_S(t)}^i = x_{G_S(t)}^i$, respectively. The process of total wheat and sorghum dry matters and the total grain dry matters  are separated because of measuring grain yield in spite of that the total wheat and sorghum dry matter contain the total grain dry matters (i.e. $X_{G_W(t)}^i \leq X_{W(t)}^i$ and $X_{G_S(t)}^i \leq X_{S(t)}^i$). Notice that since there is no sorghum crop in the Tarlee dataset, we do not consider process and measurement models of the sorghum in its model. In addition, as there is no pasture in the Brigalow dataset, we do not consider process and measurement models of the pasture in the model. Therefore, the mass of carbon inputs, $I_{C(t)}^i$, of the Tarlee and Brigalow datasets which are denoted by $IT_{C(t)}^i$ and $IB_{C(t)}^i$ respectively are:
\begin{align*}
    IT_{C(t)}^i = 
\begin{cases}
    c(X_{W(t)}^i-X_{G_W(t)}^i) + c r_W X_{W(t)}^i       & \quad \text{Wheat for Grain }\\
    c p X_{W(t)}^i + c r_W X_{W(t)}^i    & \quad \text{Wheat for Hay }\\
    c X_{P(t)}^i + c r_P X_{P(t)}^i     & \quad \text{Pasture}\\
    c p X_{P(t)}^i + c r_P X_{P(t)}^i   & \quad \text{Pasture for Hay}\\
    0                                & \quad \text{Fallow}
  \end{cases}
\end{align*}
and
\begin{align*}
    IB_{C(t)}^i = 
\begin{cases}
    c(X_{W(t)}^i-X_{G_W(t)}^i) + c r_W X_{W(t)}^i       & \quad \text{Wheat for Grain }\\
    c p X_{W(t)}^i + c r_W X_{W(t)}^i    & \quad \text{Wheat for Hay }\\
    c(X_{S(t)}^i-X_{G_S(t)}^i) + c r_S X_{S(t)}^i       & \quad \text{Sorghum for Grain }\\
    c p X_{S(t)}^i + c r_S X_{S(t)}^i    & \quad \text{Sorghum for Hay }\\
    0                                & \quad \text{Fallow}
  \end{cases}
\end{align*}
\noindent where $p$, $r_W$, $r_S$, and $r_P$ are the proportion of the crop left above-ground after harvest, the root-to-shoot ratios (in terms of TDM) for wheat, sorghum and pasture crops, respectively. The amount of carbon that enters into the soil from plant-matter that is already below-ground (i.e., roots) and from plant-matter that remains above-ground after harvesting, $c(X_{W(t)}^i-X_{G_W(t)}^i)$ and $c(X_{S(t)}^i-X_{G_S(t)}^i)$, are included in $IT_{C(t)}^i$ and $IB_{C(t)}^i$, respectively. Here $c$, $(X_{S(t)}^i-X_{G_S(t)}^i)$  and $(X_{W(t)}^i-X_{G_W(t)}^i)$ are the carbon content of dry plant matter and the above-ground plant-matter biomass of sorghum and wheat, respectively.

In this model, $X_{(t)}^i= ( X_{C(t)}^i,X_{G_W(t)}^i,X_{W(t)}^i,X_{P(t)}^i,X_{G_S(t)}^i,X_{S(t)}^i)$ is all processes at time $t$ in field (soil type) $i$. Given a vector of parameters $\boldsymbol{\theta}$ for the model, the transition density of the joint process model (which is a discrete-time Markov chain) of the independent processes (\ref{eq1S})-(\ref{eq6sS}) can be written as:
\begin{align*}
    p(X_{(t)}^i \vert X_{(t-1)}^i, \boldsymbol{\theta}) &= p(X_{C(t)}^i \vert I_{C(t)}^i, X_{C(t-1)}^i, \boldsymbol{\theta})\times p(X_{G_W(t)}^i \vert X_{G_W(t-1)}^i, \boldsymbol{\theta})\\
    &\times p(X_{W(t)}^i \vert X_{G_W(t)}^i, \boldsymbol{\theta})\times p(X_{P(t)}^i \vert X_{P(t-1)}^i, \boldsymbol{\theta})\times p(X_{G_S(t)}^i \vert X_{G_S(t-1)}^i, \boldsymbol{\theta})\\
    &\times p(X_{S(t)}^i \vert X_{G_S(t)}^i, \boldsymbol{\theta}).
\end{align*}

\noindent The overall transition density is 
\begin{align*}
    p(X_{(t)} \vert X_{(t-1)}, \boldsymbol{\theta}) = \prod _{i =1}^3 p(X_{(t)}^i \vert X_{(t-1)}^i, \boldsymbol{\theta});
\end{align*}
as the three  fields (soil types) are independent. For the sake of simplicity, we use the same notation for the overall transition density in all models. 

\subsection{Observation Model}\label{ObsOnePool}
The observation model of the one-pool includes the sub-models (\ref{eq:5})-(\ref{eq:10S}) to account for measurement error.
\begin{align}
    &Y_{TOC(t)}^i \vert X_{C(t)}^i= x_{C(t)}^i \sim LN(\log(x_{C(t)}^i), \sigma_{\epsilon TOC}^2); \label{eq:5}\\
    &Y_{G_W(t)}^i \vert X_{G_W(t)}^i= x_{G_W(t)}^i \sim LN(\log(x_{G_W(t)}^i), \sigma_{\epsilon G_W}^2); \label{eq:6}\\
    &Y_{W(t)}^i \vert X_{W(t)}^i= x_{W(t)}^i \sim LN(\log(x_{W(t)}^i), \sigma_{\epsilon W}^2); \label{eq:7}\\
    &Y_{P(t)}^i \vert X_{P(t)}^i= x_{P(t)}^i \sim LN(\log(x_{P(t)}^i), \sigma_{\epsilon P}^2); \label{eq:8} \\
    &Y_{G_S(t)}^i \vert X_{G_S(t)}^i= x_{G_S(t)}^i \sim LN(\log(x_{G_S(t)}^i), \sigma_{\epsilon G_S}^2); \quad \text{and} \label{eq:9S}\\
    &Y_{S(t)}^i \vert X_{S(t)}^i= x_{S(t)}^i \sim LN(\log(x_{S(t)}^i), \sigma_{\epsilon S}^2). \label{eq:10S}
\end{align}
As the measurements (\ref{eq:5})-(\ref{eq:10S}) are independent, the joint observation model of them at time $t$ and field (or soil type) $i$, conditioning on unobserved variable vector $X_{(t)}^i$ and parameter $\boldsymbol{\theta}$ is:
\begin{align}\label{JointObs1pool}
    p(Y_{(t)}^i \vert X_{(t)}^i, \boldsymbol{\theta}) &= p(Y_{TOC(t)}^i \vert X_{C(t)}^i, \boldsymbol{\theta}) \times p(Y_{G_W(t)}^i \vert X_{G_W(t)}^i, \boldsymbol{\theta}) \nonumber \\ 
    &\times p(Y_{W(t)}^i \vert X_{W(t)}^i, \boldsymbol{\theta}) \times p(Y_{P(t)}^i \vert X_{P(t)}^i, \boldsymbol{\theta}) \nonumber \\
    &\times p(Y_{G_S(t)}^i \vert X_{G_S(t)}^i, \boldsymbol{\theta}) \times p(Y_{S(t)}^i \vert X_{S(t)}^i, \boldsymbol{\theta});
\end{align} 
where $Y_{(t)}^i = (Y_{TOC(t)}^i, Y_{G_W(t)}^i, Y_{W(t)}^i, Y_{P(t)}^i, Y_{G_S(t)}^i, Y_{S(t)}^i)$. The overall observation model across all $i$'s is therefore:
\begin{align*}
    p(Y_{(t)} \vert X_{(t)}, \boldsymbol{\theta}) = \prod _{i =1}^3 p(Y_{(t)}^i \vert X_{(t)}^i, \boldsymbol{\theta}).
\end{align*}
For notational convenience, we use the same notation $Y_{(t)}$ and $X_{(t)}$ in other models presented below.

\section{Two-pool Model}\label{SupplTwoProcessModel}

\subsection{Process Model}\label{TwoProcessModel}
The process model of the two-pool model includes the following sub-models
\begin{align}\label{IOMprocessModel}
    &\log (X_{C(t)}^i) = \log (X_{C(t-1)}^i e^{-K_C\Delta t} + I_{C(t)}^i) + \eta _{(t)}^i, \ \ \ \eta_{(t)}^i \sim N(0, \sigma_{\eta} ^2);\nonumber\\ 
    &X_{G_W(t)}^i \sim LN(\mu _{G_W} + \rho _{G_W} (\log (X_{G_W(t-1)}^i)- \mu _{G_W}), \sigma_{G_W}^2);\nonumber \\
    &X_{W(t)}^i \sim LN(\log h_W + \log (x_{G_W(t)}^i), \sigma_W^2);\nonumber      \\
    &X_{P(t)}^i \sim LN(\mu _P + \rho _P (\log (X_{P(t-1)}^i)- \mu _P), \sigma_P^2);\nonumber \\
    &X_{G_S(t)}^i \sim LN(\mu _{G_S} + \rho _{G_S} (\log (X_{G_S(t-1)}^i)- \mu _{G_S}), \sigma_{G_S}^2);\nonumber  \\
    &X_{S(t)}^i \sim LN(\log h_S + \log (x_{G_S(t)}^i), \sigma_S^2);\nonumber \quad \text{and}\\
    &X^i_{IOM(t)} = X^i_{IOM(t-1)} = M;
\end{align}
where $M$ is an unknown constant value.

\subsection{Observation Model}\label{TwoObsModel}
The observation model of the two-pool model includes the following sub-models  
\begin{align}
    &Y_{TOC(t)}^i \vert X_{C(t)}^i= x_{C(t)}^i, X_{IOM(t)}^i= x_{IOM(t)}^i \sim LN(\log(x_{C(t)}^i + x_{IOM(t)}^i), \sigma_{\epsilon TOC}^2); \label{2PoolObs1}\\
    &Y_{IOM(t)}^i \vert X_{IOM(t)}^i= x_{IOM(t)}^i \sim LN(\log(x_{IOM(t)}^i), \sigma_{\epsilon IOM}^2); \label{2PoolObs2}\\
    &Y_{G_W(t)}^i \vert X_{G_W(t)}^i= x_{G_W(t)}^i \sim LN(\log(x_{G_W(t)}^i), \sigma_{\epsilon G_W}^2); \nonumber\\
    &Y_{W(t)}^i \vert X_{W(t)}^i= x_{W(t)}^i \sim LN(\log(x_{W(t)}^i), \sigma_{\epsilon W}^2); \nonumber\\
    &Y_{P(t)}^i \vert X_{P(t)}^i= x_{P(t)}^i \sim LN(\log(x_{P(t)}^i), \sigma_{\epsilon P}^2);  \nonumber\\
    &Y_{G_S(t)}^i \vert X_{G_S(t)}^i= x_{G_S(t)}^i \sim LN(\log(x_{G_S(t)}^i), \sigma_{\epsilon G_S}^2); \quad \text{and} \nonumber\\
    &Y_{S(t)}^i \vert X_{S(t)}^i= x_{S(t)}^i \sim LN(\log(x_{S(t)}^i), \sigma_{\epsilon S}^2).\nonumber
\end{align}
As it is shown in (\ref{2PoolObs1}), the observational data TOC depends on the state variables $X_{C(t)}^i$ and $X_{IOM(t)}^i$. Since the measurements (\ref{2PoolObs1}) and (\ref{2PoolObs2}) are independent, we compute the joint observation model of the components of this model by multiplying the probability density functions of measurement variables $Y_{TOC(t)}^i, Y_{G_W(t)}^i, Y_{W(t)}^i, Y_{P(t)}^i, Y_{G_S(t)}^i, Y_{S(t)}^i$, and $Y_{IOM(t)}^i$ given their corresponding state variables at time $t$.

\section{Three-pool Model}\label{SupplThreeProcessModel}

\subsection{Process Model}\label{1SupplThreeProcessModel}
The first pool of this model encompasses components DPM, POC, and HUM. To simplify notation, we keep the latent variable of this pool as $X_{C(t)}^i$ at time $t$ in field (soil type) $i$. The process model of the three-pool model includes the latent variables $X_{B(t)}^i$ and $X_{C(t)}^i$ along with the latent variables of processes (\ref{IOMprocessModel}) and (\ref{eq2S})-(\ref{eq6sS}). The processes of $X_{C(t)}^i$ and $X_{B(t)}^i$ are shown as
\begin{align}
    \log (X_{C(t)}^i) = &\log (X_{C(t-1)}^i e^{-K_C\Delta t} + I_{C(t)}^i \nonumber \\
    &+ X_{B(t-1)}^i(1 - e^{-K_B\Delta t})\pi _{BC}) + \eta _{C(t)}^i, \ \ \ \eta_{C(t)}^i \sim N(0, \sigma_{\eta_C} ^2);\\
    \log (X_{B(t)}^i) = &\log (X_{B(t-1)}^i e^{-K_B\Delta t} + X_{C(t-1)}^i(1 - e^{-K_C\Delta t})\pi _{CB} \nonumber \\
    &+ X_{B(t-1)}^i(1 - e^{-K_B\Delta t})\pi _{BB}) + \eta _{B(t)}^i, \ \ \ \eta_{B(t)}^i \sim N(0, \sigma_{\eta_B} ^2).
\end{align}

Notice that the mass of BIO should be less than $5$ percent of the total mass of carbon (e.g. $X_{BIO} \leq 0.05 X_{C}$). The transition density of the joint process model of the latent variables in the three-pool model is the product of the transition densities of all state variables in the three-pool model.

\subsection{Observation Model}\label{1SupplThreeObsModel}
The BIO pool is considered as the third pool in the three-pool model and uncertainty around the observations of the total SOC in field (soil type) $i$ at time $t$ is modeled as follows
\begin{align}
    Y_{TOC(t)}^i \vert X_{C(t)}^i=& ~x_{C(t)}^i, X_{IOM(t)}^i= x_{IOM(t)}^i\nonumber \\
    &, X_{B(t)}^i= x_{B(t)}^i \sim LN(\log(x_{C(t)}^i + x_{IOM(t)}^i + x_{B(t)}^i), \sigma_{\epsilon TOC}^2). \label{3PoolObs}
\end{align}
In addition to sub-model (\ref{3PoolObs}), the observation model of the three-pool model includes sub-models (\ref{eq:6})-(\ref{eq:10S}) and (\ref{2PoolObs2}). The joint observation model of the three-pool model can be computed by multiplying the probability density functions of  equations (\ref{eq:6})-(\ref{eq:10S}), (\ref{2PoolObs2}), and (\ref{3PoolObs}). The next section introduces the process and observation models of the five-pool model in which each SOC component is considered as a pool.

\section{Five-pool Model}\label{SupplFiveProcessModel}
\subsection{Process Model}
The process model of the five-pool model includes sub-models (\ref{IOMprocessModel}), (\ref{eq2S})-(\ref{eq6sS}), and the following sub-models
\begin{align}
    \log (X_{D(t)}^i) = &\log (X_{D(t-1)}^i e^{-K_D\Delta t} + P_D I_{C(t)}^i) + \eta _{D(t)}^i, \ \ \ \eta_{D(t)}^i \sim N(0, \sigma_{\eta_D} ^2); \label{F_P_M1S}\\
    \log (X_{R(t)}^i) = &\log (X_{R(t-1)}^i e^{-K_R\Delta t} + (1 - P_D) I_{C(t)}^i) + \eta _{R(t)}^i, \ \ \ \eta_{R(t)}^i \sim N(0, \sigma_{\eta_R} ^2);\\
    \log (X_{H(t)}^i) = &\log (X_{H(t-1)}^i e^{-K_H\Delta t} + X_{D(t-1)}^i(1 - e^{-K_D\Delta t})\pi _{DH}\nonumber \\
    & + X_{R(t-1)}^i(1 - e^{-K_R\Delta t})\pi _{RH}  + X_{H(t-1)}^i(1 - e^{-K_H\Delta t})\pi _{HH}\nonumber \\
    & + X_{B(t-1)}^i(1 - e^{-K_B\Delta t})\pi _{BH})  + \eta _{H(t)}^i, \ \ \ \eta_{H(t)}^i \sim N(0, \sigma_{\eta_H} ^2);\\
    \log (X_{B(t)}^i) = &\log (X_{B(t-1)}^i e^{-K_B\Delta t} + X_{D(t-1)}^i(1 - e^{-K_D\Delta t})\pi _{DB}\nonumber \\
    & + X_{R(t-1)}^i(1 - e^{-K_R\Delta t})\pi _{RB}  + X_{H(t-1)}^i(1 - e^{-K_H\Delta t})\pi _{HB}\nonumber \\
    & + X_{B(t-1)}^i(1 - e^{-K_B\Delta t})\pi _{BB})  + \eta _{B(t)}^i, \ \ \ \eta_{B(t)}^i \sim N(0, \sigma_{\eta_B} ^2). \label{F_P_M4S}
\end{align}

The transition density of the joint process model of the latent variables in this model can be gained through multiplying the transition densities of the latent variables in this model as they are independent.

\subsection{Observation Model}
The observation model of the five-pool model captures the uncertainties in the observations of the carbon input, TOC, POC, HUM, and IOM. The observation models of the carbon input and IOM which are presented by equations (\ref{eq:6})-(\ref{eq:10S}) and (\ref{2PoolObs2}) are the same in this model. The measurement processes of the TOC, POC, and HUM given their related state variables (e.g. $X_{D(t)}^i = x_{D(t)}^i$) are
\begin{align}
    \log(Y_{TOC(t)}^i) = \log&(x_{D(t)}^i + x_{IOM(t)}^i + x_{B(t)}^i + x_{R(t)}^i + x_{H(t)}^i) +\eta_{\epsilon TOC}, \ \ \ \eta_{\epsilon TOC} \sim N(0, \sigma_{\epsilon TOC}^2); \label{5PoolObs1} \\
     \log(Y_{POC(t)}^i) = \log&(x_{D(t)}^i + x_{B(t)}^i + x_{R(t)}^i) +\eta_{\epsilon POC}, \ \ \ \eta_{\epsilon POC} \sim N(0, \sigma_{\epsilon POC}^2); \label{5PoolObs2} \\
     \log(Y_{H(t)}^i) = \log(x&_{H(t)}^i) +\eta_{\epsilon H}, \ \ \ \eta_{\epsilon H} \sim N(0, \sigma_{\epsilon H}^2). \label{5PoolObs3} 
\end{align}
\noindent Since the measurement variables in (\ref{5PoolObs1})-(\ref{5PoolObs3}) are independent, the product of the probability density functions of  equations (\ref{eq:6})-(\ref{eq:10S}), (\ref{2PoolObs2}), and (\ref{5PoolObs1})-(\ref{5PoolObs3}) yields the joint observation model of the five-pool model.

\section{Gelman and Rubin’s Convergence Diagnostic Statistic}\label{GelmanAndRubin}
\begin{table}[ht]
\begin{center}
\begin{tabular}{|c|c|c|}
\hline
 \cellcolor{gray!60} Parameter & \cellcolor{gray!60} $\hat{R}$ & \cellcolor{gray!60} Upper C.I. bound on $\hat{R}$ \\
\hline
$K_C$ & 1.00 & 1.00 \\
\hline
$c$ & 1.00 & 1.00 \\
\hline
$r_W$ & 1.00 & 1.00 \\
\hline
$r_P$ & 1.00 & 1.00 \\
\hline
p & 1.00 & 1.00 \\
\hline
$h_W$ & 1.00 & 1.01 \\
\hline
$\mu_{G_W}$ & 1.03 & 1.08 \\
\hline
$\mu_P$ & 1.00 & 1.00 \\
\hline
$\rho_{G_W}$ & 1.06 & 1.16 \\
\hline
$\rho_P$ & 1.00 & 1.00 \\
\hline
$\sigma_{\eta_C}^2$ & 1.05 & 1.07 \\
\hline
$\sigma_{G_W}^2$ & 1.01 & 1.02 \\
\hline
$\sigma_{W}^2$ & 1.07 & 1.11 \\
\hline
$\sigma_{P}^2$ & 1.01 & 1.01 \\
\hline 
$Y_{IOM}$ & 1.00 & 1.00 \\
\hline 
$\sigma_{\eta_B}^2$ & 1.02 & 1.04 \\
\hline 
$K_{B}$ & 1.00 & 1.00 \\
\hline 
$\pi_{CB}$ & 1.00 & 1.01 \\
\hline 
$\pi_{BB}$ & 1.00 & 1.00 \\
\hline$
\pi_{BC}$ & 1.00 & 1.00 \\
\hline
\end{tabular}
\caption{\label{diag} The Gelman and Rubin's convergence diagnostic, $\hat{R}$ calculated for model parameters of the three-pool model of the Tarlee dataset. Since the point estimate of $\hat{R}$ for each parameter is less than 1.2, the MCMC samples can be considered to have reached a stationary distribution and are mixing adequately.}
\end{center}
\end{table}

\begin{table}[ht]
\begin{center}
\begin{tabular}{|c|c|c|}
\hline
 \cellcolor{gray!60} Parameter & \cellcolor{gray!60} $\hat{R}$ & \cellcolor{gray!60} Upper C.I. bound on $\hat{R}$ \\
\hline
$K_C$ & 1.00 & 1.00 \\
\hline
$c$ & 1.00 & 1.00 \\
\hline
$r_W$ & 1.00 & 1.00 \\
\hline
p & 1.01 & 1.03 \\
\hline
$h_W$ & 1.01 & 1.02 \\
\hline
$\mu_{G_W}$ & 1.10 & 1.23 \\
\hline
$\mu_{G_S}$ & 1.00 & 1.00 \\
\hline
$\rho_{G_W}$ & 1.01 & 1.03 \\
\hline
$\sigma_{\eta}^2$ & 1.07 & 1.19 \\
\hline
$\sigma_{G_W}^2$ & 1.00 & 1.00 \\
\hline
$\sigma_{W}^2$ & 1.01 & 1.01 \\
\hline
$r_{S}$ & 1.00 & 1.00 \\
\hline
$\sigma_{B}^2$ & 1.14 & 1.35 \\
\hline
$\sigma_{S}^2$ & 1.10 & 1.30 \\
\hline
$K_{B}$ & 1.00 & 1.02 \\
\hline
$K_{R}$ & 1.02 & 1.07 \\
\hline
$\pi_{DB}$ & 1.00 & 1.00 \\
\hline
$\pi_{BB}$ & 1.00 & 1.00 \\
\hline
$\pi_{BC}$ & 1.00 & 1.02 \\
\hline
$\pi_{CB}$ & 1.05 & 1.17 \\
\hline
$\mu_{G_S}$ & 1.00 & 1.00 \\
\hline
$\rho_{G_S}$ & 1.00 & 1.00 \\
\hline
$\sigma_{G_S}^2$ & 1.05 & 1.07 \\
\hline
$h_S$ & 1.01 & 1.02 \\
\hline
$\rho_{G_S}$ & 1.00 & 1.00 \\
\hline
\end{tabular}
\caption{\label{diagBrigalow} The Gelman and Rubin's convergence diagnostic, $\hat{R}$ calculated for model parameters of the three-pool model of the Brigalow dataset. Since the point estimate of $\hat{R}$ for each parameter is less than 1.2, the MCMC samples can be considered to have reached a stationary distribution and are mixing adequately.}
\end{center}
\end{table}

\section{Estimated LPD and ELPD of the Models}\label{LPDSection}
The mean and standard deviation of the fourchains of the estimated LPD at each time point and the ELPD of the models applied on the Tarleeand Brigalow datasets are shown in Tables \ref{LFOTable} and \ref{LFOBrigTable}, respectively.

\begin{table}[!ht]
\begin{center}
\begin{tabular}{|l|l|l|l|l|l|l|l|l|}
\hline
\rowcolor[HTML]{9B9B9B} 
\multicolumn{1}{|c|}{\cellcolor[HTML]{9B9B9B}}                       & \multicolumn{2}{c|}{\cellcolor[HTML]{9B9B9B}One-pool model}                                          & \multicolumn{2}{c|}{\cellcolor[HTML]{9B9B9B}Two-pool model}                                                                 & \multicolumn{2}{c|}{\cellcolor[HTML]{9B9B9B}Three-pool model}                                        & \multicolumn{2}{c|}{\cellcolor[HTML]{9B9B9B}Five-pool model}                                         \\  \cline{2-9}
\rowcolor[HTML]{9B9B9B} 
\multicolumn{1}{|c|}{\cellcolor[HTML]{9B9B9B}} & 
\multicolumn{1}{c|}{\cellcolor[HTML]{9B9B9B}Mean} &
\multicolumn{1}{c|}{\cellcolor[HTML]{9B9B9B}SD} & 
\multicolumn{1}{c|}{\cellcolor[HTML]{9B9B9B}Mean} & 
\multicolumn{1}{c|}{\cellcolor[HTML]{9B9B9B}SD} & \multicolumn{1}{c|}{\cellcolor[HTML]{9B9B9B}Mean} & 
\multicolumn{1}{c|}{\cellcolor[HTML]{9B9B9B}SD} & 
\multicolumn{1}{c|}{\cellcolor[HTML]{9B9B9B}Mean} & 
\multicolumn{1}{c|}{\cellcolor[HTML]{9B9B9B}SD} 
 \\  
\rowcolor[HTML]{9B9B9B} 
\multicolumn{1}{|c|}{\multirow{-3}{*}{\cellcolor[HTML]{9B9B9B}Time}} &
\multicolumn{1}{c|}{\cellcolor[HTML]{9B9B9B}(LPD)} &
\multicolumn{1}{c|}{\cellcolor[HTML]{9B9B9B}(LPD)} &
\multicolumn{1}{|c|}{\cellcolor[HTML]{9B9B9B}(LPD)} & 
\multicolumn{1}{c|}{\cellcolor[HTML]{9B9B9B}(LPD)} &
\multicolumn{1}{c|}{\cellcolor[HTML]{9B9B9B}(LPD)} &
\multicolumn{1}{|c|}{\cellcolor[HTML]{9B9B9B}(LPD)} & 
\multicolumn{1}{c|}{\cellcolor[HTML]{9B9B9B}(LPD)} &
\multicolumn{1}{|c|}{\cellcolor[HTML]{9B9B9B}(LPD)} 
\\ \cline{3-9}
\hline
13 & -5.27  & 0.44  &  -3.21 & 0.27 & -2.40 & 0.04 & -2.67 & 0.02\\ 
\hline
14 & -8.14  & 0.41  & -8.34 & 0.48 & -7.22 & 0.17 & -6.35 & 0.10\\ 
\hline
15 & -5.53  & 0.30  & -3.72 & 0.08 & -3.26 & 0.04 & -2.85 & 0.01\\ 
\hline
16 & -7.82  & 0.41  & -7.32 & 0.77 & -6.09 & 0.24 & -6.09 & 0.27\\ 
\hline
17 & -6.45  & 0.86  & -5.15 & 0.65 & -5.34 & 0.36 & -3.62 & 0.02\\ 
\hline
18 & -7.70  & 0.50 & -6.42 & 0.21 & -5.11 & 0.21 & -4.97 & 0.12\\ 
\hline
19 & -5.66 & 0.44 & -3.02 & 0.11 & -2.59 & 0.03 & -3.04 & 0.03\\ 
\hline
20 & -6.45 & 0.44 & -3.33 & 0.25 & -2.78 & 0.71 & -7.41 & 0.61 \\ 
\hline
ELPD & -53.02 & 3.80   & -40.55 & 2.82  &\textbf{\textcolor{red}{-34.79}} & 1.80 & -37 & 1.18 \\ 
\hline
\end{tabular}
\end{center}
\caption{The mean and standard deviation (SD) of the four chains of the estimated LPD and ELPD of the SOC models applied on the Tarlee dataset.}
\label{LFOTable}
\end{table}

\begin{table}[!ht]
\begin{center}
\begin{tabular}{|l|l|l|l|l|l|l|l|l|}
\hline
\rowcolor[HTML]{9B9B9B} 
\multicolumn{1}{|c|}{\cellcolor[HTML]{9B9B9B}}                       & \multicolumn{2}{c|}{\cellcolor[HTML]{9B9B9B}One-pool model}                                          & \multicolumn{2}{c|}{\cellcolor[HTML]{9B9B9B}Two-pool model}                                                                 & \multicolumn{2}{c|}{\cellcolor[HTML]{9B9B9B}Three-pool model}                                        & \multicolumn{2}{c|}{\cellcolor[HTML]{9B9B9B}Five-pool model}                                         \\  \cline{2-9}
\rowcolor[HTML]{9B9B9B} 
\multicolumn{1}{|c|}{\cellcolor[HTML]{9B9B9B}} & 
\multicolumn{1}{c|}{\cellcolor[HTML]{9B9B9B}Mean} &
\multicolumn{1}{c|}{\cellcolor[HTML]{9B9B9B}SD} & 
\multicolumn{1}{c|}{\cellcolor[HTML]{9B9B9B}Mean} & 
\multicolumn{1}{c|}{\cellcolor[HTML]{9B9B9B}SD} & \multicolumn{1}{c|}{\cellcolor[HTML]{9B9B9B}Mean} & 
\multicolumn{1}{c|}{\cellcolor[HTML]{9B9B9B}SD} & 
\multicolumn{1}{c|}{\cellcolor[HTML]{9B9B9B}Mean} & 
\multicolumn{1}{c|}{\cellcolor[HTML]{9B9B9B}SD} 
 \\  
\rowcolor[HTML]{9B9B9B} 
\multicolumn{1}{|c|}{\multirow{-3}{*}{\cellcolor[HTML]{9B9B9B}Time}} &
\multicolumn{1}{c|}{\cellcolor[HTML]{9B9B9B}(LPD)} &
\multicolumn{1}{c|}{\cellcolor[HTML]{9B9B9B}(LPD)} &
\multicolumn{1}{|c|}{\cellcolor[HTML]{9B9B9B}(LPD)} & 
\multicolumn{1}{c|}{\cellcolor[HTML]{9B9B9B}(LPD)} &
\multicolumn{1}{c|}{\cellcolor[HTML]{9B9B9B}(LPD)} &
\multicolumn{1}{|c|}{\cellcolor[HTML]{9B9B9B}(LPD)} & 
\multicolumn{1}{c|}{\cellcolor[HTML]{9B9B9B}(LPD)} &
\multicolumn{1}{|c|}{\cellcolor[HTML]{9B9B9B}(LPD)} 
\\ \cline{3-9}
\hline
14 & -9.82 & 0.82 & -9.93 & 1.70 & -8.62 & 0.45 & -9.60 & 1.71 \\ 
\hline
15 & -7.13 & 0.22 & -6.88 & 0.14 & -6.97 & 0.34 & -7.40 & 0.43 \\ 
\hline
16 & -7.86 & 0.50 & -7.96 & 0.87 & -8.51 & 0.47 & -8.62 & 0.54 \\ 
\hline
17 & -3.69 & 0.06 & -3.71 & 0.05 & -3.77 & 0.06 & -3.94 & 0.03 \\ 
\hline
18 & -6.84 & 0.28 & -6.74 & 0.05 & -6.57 & 0.56 & -7.95 & 0.98 \\ 
\hline
19 & -1.55 & 0.03 & -1.66 & 0.03 & -2.04 & 0.06 & -12.06  & 0.89 \\ 
\hline
ELPD & -36.89 & 1.91   & -36.88 & 2.84  &\textbf{\textcolor{red}{-36.48}} & 1.94 & -49.57 & 4.58 \\ 
\hline
\end{tabular}
\end{center}
\caption{The mean and standard deviation (SD) of the four chains of the estimated LPD and ELPD of the SOC models applied on the Brigalow dataset.}
\label{LFOBrigTable}
\end{table}


\begin{figure}[ht]
    \centering
    \includegraphics[width=15cm, height=12.2cm]{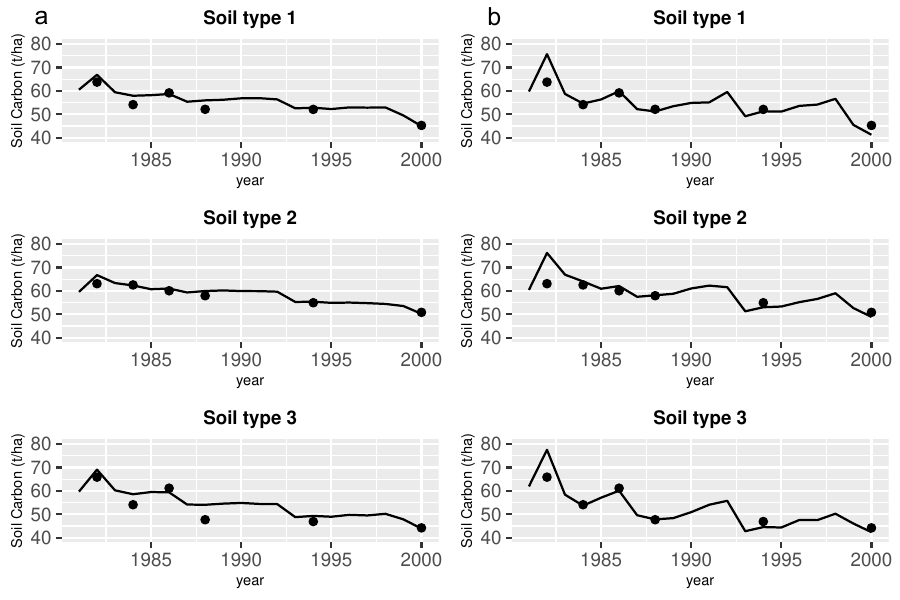}
    \vspace*{-0.5cm}
    \caption{Soil organic carbon (SOC) dynamics of the Brigalow dataset based on a) the three-pool model and b) the five-pool model. The $50^{th}$ percentile is shown by the solid line and the measured SOC values are indicated by filled dots.}
    \label{50PercentBrigalow}
\end{figure}

\section{Hardware Use and Computing Time}\label{Sec:Elapsedtime}
The codes are run on High Performance
Computing (HPC) in R version 3.5.1. The elapsed running time, per minute, of the codes per $1000$ MCMC iterations is provided in Table \ref{ElapseTime}.

\begin{table}[ht]
\begin{center}
\begin{tabular}{|
>{\columncolor[HTML]{9B9B9B}}l |c|c|}
\hline
\multicolumn{1}{|c|}{\cellcolor[HTML]{9B9B9B}}                        & \multicolumn{2}{c|}{\cellcolor[HTML]{9B9B9B}Time (Mins)}          \\ \cline{2-3} 
\multicolumn{1}{|c|}{\multirow{-2}{*}{\cellcolor[HTML]{9B9B9B}Model}} & \cellcolor[HTML]{9B9B9B}Brigalow & \cellcolor[HTML]{9B9B9B}Tarlee \\ \hline
One-pool   & 0.83 & 2.20 \\ \hline
Two-pool   & 1.29 & 3.03 \\ \hline
Three-pool & 1.46 & 1.53 \\ \hline
Five-pool  & 1.50 & 2.07 \\ \hline
\end{tabular}
\caption{The elapsed running time (per minute) of the codes. The elapsed time is based on $1000$ MCMC iterations.}
    \label{ElapseTime}
    \end{center}
\end{table}

\end{document}